\RequirePackage{etoolbox}
\csdef{input@path}{{style/}{graphics/}}
\documentclass[aoas,MSNbibl,nameyear,seceqn,dvips]{arximspdf}
\makeatletter
   \@ifpackageloaded{graphicx}{}{\usepackage{graphicx}}
\makeatother

\doi{10.1214/14-AOAS799}
\volume{9}
\issue{1}
\pubyear{2015}
\firstpage{94}
\lastpage{121}
\docsubty{FLA}

\makeatletter
\newcommand{\gls}{\mathrm{g.l.s.}}
\makeatother

\begin{document}
\begin{frontmatter}

\title{Characterizing the spatial structure of defensive skill in
professional basketball}
\runtitle{Defensive skill in basketball}

\begin{aug}
\author[A]{\fnms{Alexander}~\snm{Franks}\corref{}\ead[label=e1]{afranks@fas.harvard.edu}},
\author[B]{\fnms{Andrew}~\snm{Miller}\ead[label=e2]{acm@seas.harvard.edu}},
\author[A]{\fnms{Luke}~\snm{Bornn}\ead[label=e3]{bornn@stat.harvard.edu}}
\and
\author[C]{\fnms{Kirk}~\snm{Goldsberry}\ead[label=e4]{kgoldsberry@fas.harvard.edu}}
\runauthor{Franks, Miller, Bornn and Goldsberry}
\affiliation{Harvard University}
\address[A]{A. Franks\\
L. Bornn\\
Department of Statistics\\
Harvard University\\
1 Oxford Street\\
Cambridge, Massachusetts 02138\\
USA\\
\printead{e1}\\
\phantom{E-mail: }\printead*{e3}}
\address[B]{A. Miller\\
Department of Computer Science\\
Harvard University\\
33 Oxford Street\\
Cambridge, Massachusetts 02138\\
USA\\
\printead{e2}}
\address[C]{K. Goldsberry\\
Institute of Quantitative Social Science\\
Harvard University\\
33 Oxford Street\\
Cambridge, Massachusetts 02138\\
USA\\
\printead{e4}}
\end{aug}

%
\received{\smonth{4} \syear{2014}}
%
\revised{\smonth{12} \syear{2014}}

%
\begin{abstract}
Although basketball is a dualistic sport, with all players competing on
both offense and defense, almost all of the sport's conventional
metrics are
designed to summarize offensive play. As a result, player valuations are
largely based on offensive performances and to a much lesser degree on
defensive ones. Steals, blocks and defensive rebounds provide only a limited
summary of defensive effectiveness, yet they persist because they summarize
salient events that are easy to observe. Due to the inefficacy of traditional
defensive statistics, the state of the art in defensive analytics remains
qualitative, based on expert intuition and analysis that can be prone
to human biases and imprecision.

Fortunately, emerging optical player tracking systems have the
potential to enable a richer quantitative characterization of
basketball performance, particularly defensive performance.
Unfortunately, due to computational and methodological complexities,
that potential remains unmet. This paper attempts to fill this void,
combining spatial and spatio-temporal processes, matrix factorization
techniques and hierarchical regression models with player tracking data to
advance the state of defensive analytics in the NBA. Our approach
detects, characterizes and
quantifies multiple aspects of defensive play in basketball, supporting
some common understandings of defensive
effectiveness, challenging others and opening up many new insights into
the defensive elements of basketball. 
\end{abstract}

%
\end{frontmatter}

\section{Introduction}\label{secintro}

In contrast to American football, where different sets of players
compete on offense and defense, in basketball every player must play
both roles. Thus, traditional ``back of the baseball card'' metrics
which focus on offensive play are inadequate for fully characterizing
player ability. Specifically, the traditional box score includes
points, assists, rebounds, steals and blocks per game, as well as
season averages like field goal percentage and free throw percentage.
These statistics paint a more complete picture of the \emph{offensive}
production of a player, while steals, blocks and defensive rebounds
provide only a limited summary of \emph{defensive} effectiveness.
These metrics, though they explain only a small fraction of defensive
play, persist because they summarize recognizable events that are
straightforward to record.

A deeper understanding of defensive skill requires that we move beyond
simple observables. Due to the inefficacy of traditional defensive
statistics, modern understanding of defensive skill has centered around
expert intuition and analysis that can be prone to human biases and
imprecision. In general, there has been little research characterizing
individual player habits in dynamic, goal-based sports such as
basketball. This is due to: (1) the lack of relevant data, (2) the
unique spatial-temporal nature of the sport, and (3) challenges
associated with disentangling confounded player effects.

One of the most popular metrics for assessing player ability,
individual plus/minus, integrates out the details of play, focusing
instead on aggregate outcomes. This statistic measures the total team
point or goal differential while a player is in the game. As such, it
represents a notion of overall skill that incorporates both offensive
and defensive ability. The biggest difficulty with individual
plus/minus, however, is player confounding. That is, plus/minus depends
crucially on the skill of an individual's teammates. One solution to
this problem is to aggregate the data further by recording empirical
plus/minus for all pairs or even triplets of players in the game
[\citet
{kubatko}]. As an alternative, several approaches control for
confounding using regression adjusted methods [\citet
{rosenbaum,sill2010improved,macdonald2011regression}]. 

Only recently have more advanced hierarchical models been used to
analyze individual player ability in sports. 
In hockey, for instance, competing process hazard models have been used
to value players, whereby outcomes are goals, with censoring occurring
at each player change [\citet{hockey}]. As with all of the plus/minus
approaches discussed earlier, this analysis looked at discrete
outcomes, without taking into consideration within-possession events
such as movements, passes and spatial play formations. Without
analyzing the spatial actions occurring within a possession, measuring
individual traits as separate from team characteristics is fraught with
identifiability problems.

There is an emerging solution to these identifiability concerns,
however, as player tracking systems become increasingly prevalent in
professional sports arenas.
{While the methodology developed herein applies to basketball on all
continents,} for this research we use optical player tracking data from
the 2013--2014 NBA season. The data, which is derived from cameras
mounted in stadium rafters, consist primarily of $x,y$ coordinates for
the ball and all ten athletes on the court (five on each team),
recorded at $25$ frames per second. In addition, the data include game
and player specific annotations: who possesses the ball, when fouls
occur and shot outcomes.

This data enables us for the first time to use spatial and
spatio-temporal information to solve some of the challenges associated
with individual player analysis. The spatial resolution of these data
have changed the types of questions we can answer about the game,
allowing for in-depth analyses into individual players [\citeauthor{goldsberry2012}
(\citeyear{goldsberry2012,goldsberry2013})]. Model-based approaches using this
rich data have also recently gained traction, with
\citet{cervone} employing multi-scale semi-Markov models to
conduct real-time evaluations of basketball plays.

While it is clear that player tracking systems have the potential to
enable a richer quantitative characterization of basketball
performance, this potential has not yet been met, particularly for
measuring defensive performance. Rather than integrate out the details
of play, we exploit the spatio-temporal information in the data to
learn the circumstances that lead to a particular outcome. In this way,
we infer not just who benefits their team, but \emph{why} and \emph
{how} they do so. Specifically, we develop a model of the spatial
behavior of NBA basketball players which reveals interpretable
dimensions of both offensive and defensive efficacy. We suspect the
proposed methodology might also find use in other sports.

\subsection{Method overview}

We seek to fill a void in basketball analytics by providing the first
quantitative characterization of man-to-man defensive effectiveness in
different regions of the court. To this end, we propose a model which
explains both shot selection (who shoots and where) as well as the
expected outcome of the shot, given the defensive assignments. We term
these quantities shot \textit{frequency} and \textit{efficiency},
respectively; see \citet{nbaglossary} for a glossary of other
basketball terms used throughout the paper. Despite the abundance of
data, critical information for determining these defensive habits is
unavailable. First and most importantly, the defensive matchups are
unknown. While it is often clear to a human observer who is guarding
whom, such information is absent from the data. While in theory we
could use crowd-sourcing to learn who is guarding whom, annotating the
data set is a subjective and labor-intensive task. Second, in order to
provide meaningful spatial summaries of player ability, we must define
relevant court regions in a data driven way. Thus, before we can begin
modeling defensive ability, we devise methods to learn these features
from the available data.




Our results reveal other details of play that are not readily apparent.
As one example, we demonstrate that two highly regarded defensive
centers, Roy Hibbert and Dwight Howard, impact the game in opposing
ways. Hibbert reduces shot efficiency near the basket more than any
other player in the game, but also faces more shots there than similar
players. Howard, on the other hand, is one of the best at reducing shot
frequency in this area, but tends to be worse than average at reducing
shot efficiency. We synthesize the spatially varying efficiency and
frequency results visually in the \emph{defensive} shot chart, a new
analogue to the oft depicted offensive shot chart.


\section{Who's guarding whom}
\label{secguarding}

For each possession, before modeling defensive skill, we must
establish some notion of defensive intent. To this end, we first
construct a model to identify which offender is guarded by each
defender at every moment in time. To identify who's guarding whom, we
infer the canonical, or central, position for a defender guarding a particular
offender at every time $t$ as a function of space--time covariates. A
player deviates from this position due to player or team specific
tendencies and unmodeled covariates. Throughout each possession, we
index each defensive player by $j \in1, \ldots, 5$ and each offensive
player by $k \in1, \ldots, 5$. Without loss of generality, we
transform the space so that all possessions occur in the same half.
To start, we model the canonical defensive location for a defender at
time $t$, guarding offender $k$, as a convex combination of three
locations: the position of the offender, $O_{tk}$, the current
location of the ball, $B_t$, and the location of the hoop, $H$. Let
$\mu_{tk}$ be the canonical location for a defender guarding player
$k$ at time $t$. Then,
\begin{eqnarray*}
\mu_{tk} &=& \gamma_oO_{tk} +
\gamma_bB_t+\gamma_hH,
\\
\Gamma\mathbf{1} &=& 1
\end{eqnarray*}
with $\Gamma=[\gamma_o, \gamma_b, \gamma_h]$.

Let $I_{tjk}$ be an indicator for whether defender $j$ is guarding
offender $k$ at time~$t$. Multiple defenders can guard the same
offender, but each defender can only be guarding one offender at any
instant. The observed location of a
defender $j$, given that they are guarding offender $k$, is normally
distributed about the mean location
\[
D_{tj}|I_{tjk}=1 \sim N\bigl(\mu_{tk},
\sigma^2_D\bigr). %
\]

We model the evolution of man-to-man defense (as given by the matrix of
matchups, $\mathbf{I}$) over the course of a
possession using a hidden Markov model. The hidden states represent
the offender that is being guarded by each defensive player. The
complete data likelihood is
\begin{eqnarray*}
L\bigl(\Gamma,\sigma^2_D\bigr) &=& P\bigl(\mathbf{D},
\mathbf{I}|\Gamma,\sigma^2_D\bigr)
\\
&=& \prod_{t,j,k} \bigl[P\bigl(D_{tj}|I_{tjk},
\Gamma,\sigma^2_D\bigr) P(I_{tjk} |
I_{(t-1)j_\cdot})\bigr]^{I_{tjk}},
\end{eqnarray*}
where $P(D_{tj}|I_{tjk}=1,\Gamma,\sigma^2_D)$ is a normal density as
stated above.
We also assume a constant transition probability, that is, a defender is
equally likely, a priori, to switch to guarding any offender at every
instant
\begin{eqnarray*}
P(I_{tjk} = 1 | I_{(t-1)jk} = 1) &=&\rho,
\\
P(I_{tjk} = 1 | I_{(t-1)jk'} = 1) &=& \frac{1-\rho}{4},\qquad k'\neq k
\end{eqnarray*}
for all defenders, $j$. {Although in reality there should be
heterogeneity in $\rho$ across players, for computational
simplicity we assume homogeneity and later show that we still
do a good job recovering switches and who's guarding whom.}
The complete log likelihood is
\begin{eqnarray*}
\ell\bigl(\Gamma,\sigma^2_D\bigr) &=& \log P\bigl(
\mathbf{D},\mathbf{I}|\Gamma,\sigma^2_D\bigr)
\\
&=& \sum_{t,j,k} {I_{tjk}}\bigl[\log\bigl(P
\bigl(D_{tj}|I_{tjk},\Gamma,\sigma^2_D
\bigr)\bigr)+\log\bigl(P(I_{tjk} | I_{(t-1)j_\cdot })\bigr)\bigr]
\\
&=& \sum_{t,j,k} \frac{I_{tjk}}{\sigma^2_D}(D_{tj} -
\mu_{tk})^2+I_{tjk}\log P(I_{tjk} |
I_{(t-1)j_\cdot }).
\end{eqnarray*}

\subsection{Inference}

We use the EM algorithm to estimate the relevant unknowns,
$I_{tjk}, \sigma^2_D$, $\Gamma$ and $\rho$. At each iteration, $i$,
of the algorithm, we perform the E-step and M-step until convergence.
In the E-step, we compute
$E_{tjk}^{(i)} =
E[I_{tjk}|D_{tj},\hat{\Gamma}^{(i)}, \hat{\sigma}_D^{2(i)},\hat
{\rho}^{(i)}]$
and
$A^{(i)}_{tjkk'}=[I_{tjk}I_{(t-1)jk'}|D_{tj},\hat{\Gamma}^{(i)},\hat
{\sigma}_D^{2(i)},\hat{\rho}^{(i)}]$
for all $t$, $j$, $k$ and~$k'$. These expectations can be computed
using the forward--backward algorithm [\citet{bishop2006}]. Since we
assume each defender acts independently, we run the forward--backward
algorithm for each $j$, to compute the expected assignments
($E_{tjk}^{(i)}$) and the probabilities\vspace*{1pt} for every pair of two
successive defensive assignments ($A^{(i)}_{tjkk'}$) for each defender
at every moment. In the M-step, we update the maximum likelihood
estimates of $\sigma^2_D$, $\Gamma$ and $\rho$ given the current
expectations.

Let $\mathbf{X} = [\mathbf{O}, \mathbf{B}, \mathbf{H}]$ be the
design matrix
corresponding to the offensive location, ball location and hoop
location. We define $X_{tk} =[O_{tk}, B_t, H]$ to be the row of the
design matrix corresponding to offender $k$ at time $t$.

In the $i$th iteration of the M-step we first update our estimates of
$\Gamma$ and $\sigma^2_D$,
\[
\bigl(\hat{\Gamma}^{(i)},\hat{\sigma}_D^{2(i)}\bigr)
\leftarrow\mathop{\arg\max}_{\Gamma, \sigma^2_D} \sum_{t,j,k}
\frac{E_{tjk}^{(i-1)} }{\sigma^2_D}(D_{tj} -\Gamma X_{tk})^2,
\qquad\Gamma\mathbf{1}=1.
\]

This maximization corresponds to the solution of a constrained
generalized least squares problem and can be found analytically. Let
$\Omega$ be the diagonal matrix of weights, in this case whose
entries at each iteration are
${\sigma^2_D}/{E_{tjk}^{(i)}}$. As $\hat{\Gamma}$ is
the maximum likelihood estimator subject to the constraint that
$\hat{\Gamma}\mathbf{1} = 1$, it can be shown that
\[
\hat{\Gamma} = \hat{\Gamma}_{\gls} + \bigl(X^T
\Omega^{-1}X\bigr)^{-1}\mathbf{1}^T\bigl(
\mathbf{1} \bigl(X^T\Omega^{-1}X\bigr)^{-1}
\mathbf{1}^T\bigr)^{-1}(1-\hat{\Gamma}_{\gls}
\mathbf{1}),
\]
where $\hat{\Gamma}_{\gls} =
(X^T\Omega^{-1}X)^{-1}X^{T}\Omega^{-1}D$ is the usual generalized
least squares estimator. Finally, the estimated defender variation at
iteration $i$, $\hat{\sigma}^2$, is simply
\[
\hat{\sigma}_D^2 = \frac{(D-\hat{\Gamma}X)^T\mathcal{E}(D-\hat
{\Gamma}X)}{N_X}, %
\]
where $\mathcal{E}=\operatorname{diag}(E_{tjk}^{(i-1)})$ for all
$t$, $j$, $k$ in iteration $i$ and $N_X=\operatorname{nrow}(X)$.


Next, we update our estimate of the transition parameter, $\rho$, in
iteration $i$:
\[
\hat{\rho}^{(i)} \leftarrow\mathop{\arg\max}_{\rho} \sum
_{t,j,k}\sum_{k'\neq k}
A_{tjkk'}\log\biggl(\frac{1-\rho}{4} \biggr)+\sum
_{t,j,k} A_{tjkk}\log(\rho). %
\]

It is easy to show, under the proposed transition model, that the maximum
likelihood estimate for the odds of staying in the same state, $ Q =
\frac{\rho}{1-\rho}$, is
\[
\hat{Q} = \frac{1}{4}\frac{\sum_{t,j,k} A_{tjkk}}{\sum_{t,j,k}\sum
_{k'\neq k}
A_{tjkk'}}
\]
and, hence, the maximum likelihood estimate for $\rho$ is
\[
\hat{\rho} = \frac{\hat{Q}}{1+\hat{Q}}.
\]

Using the above equations, we
iterate until convergence, saving the final estimates of
$\hat{\Gamma}$, $\hat{\sigma}_D^2$ and $\hat{\rho}$.

\subsection{Results}

\begin{figure}

\includegraphics{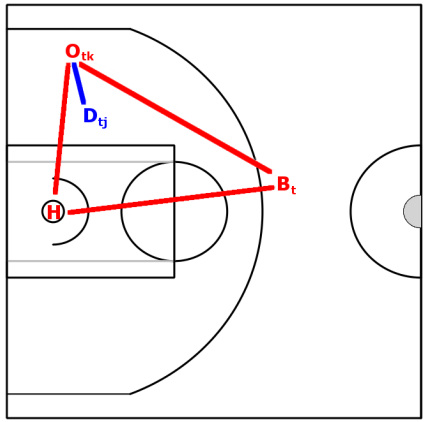}

\caption{The canonical defending location is a convex combination of
the offender, ball and hoop locations.}
\label{opt-triang}
\end{figure}

First, we restrict our analysis to the parts of a possession in which
all players are in the offensive half court---when the ball is moved
up the court at the beginning of each possession, most defenders are
not yet actively guarding an offender. We use the EM algorithm to fit
the HMM on 30 random possessions from the database. We find that a
defender's canonical position can be described as
\mbox{$0.62O_{tk}+0.11B_t+0.27H$} at any moment in time. That is, we infer that
on average the defenders position themselves just over two thirds\vspace*{1pt}
($\frac{0.62}{0.27+0.62}\approx0.70$) of the way between the hoop and
the offender they are guarding, shading slightly toward the ball (see
Figure~\ref{opt-triang}). Since the weights are defined on a
relative rather than absolute scale, the model accurately reflects
the fact that defenders guard players more closely when they are near
the basket. Furthermore, the model captures the fact that a defender
guards the ball carrier more closely, since the ball and the offender
are in roughly the same position. In this case, on average, the
defender positions himself closer to three fourths ($0.73O_{tk}+0.27H$) of
the way between the ball carrier and the basket.

%

\begin{figure}
\begin{tabular}{@{}c@{\hspace*{8pt}}c@{\hspace*{8pt}}c@{}}

\includegraphics{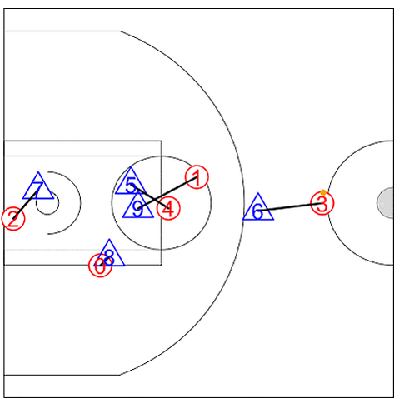}
 & \includegraphics{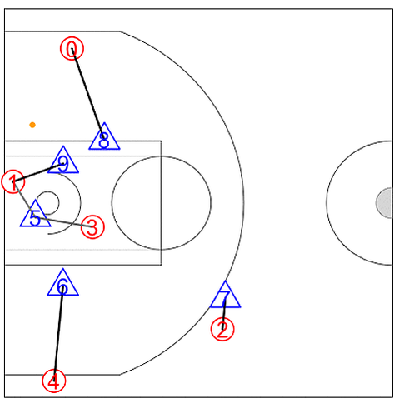} & \includegraphics{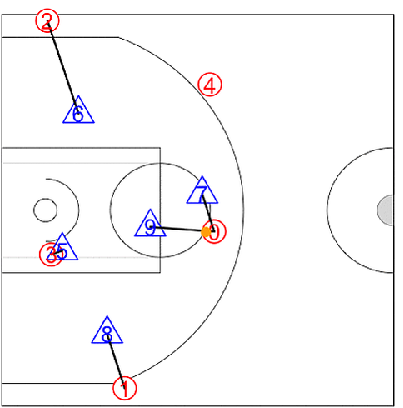}\\
\footnotesize{(a)} & \footnotesize{(b)} & \footnotesize{(c)}
\end{tabular}
\caption{Who's guarding whom. Players 0--4 (red circles) are the
offenders and players 5--9 (blue triangles) are defenders. Line
darkness represents degree of certainty. We illustrate a few
properties of the model: \textup{(a)} defensive assignments are not just
about proximity---given this snapshot, it appears as if 5 should be
guarding 1 and 9 should be guarding 4. However, from the full
animation, it is clear that 9 is actually chasing 1 across the
court. The HMM enforces some smoothness, which ensures that we
maintain the correct matchups over time. \textup{(b)} We capture uncertainty
about who is guarding whom, as illustrated by multiple faint lines
from defender 5. There is often more uncertainty near the basket.
\textup{(c)} Our model captures double teams (defenders 7 and 9 both guarding
0). Full animations are available in Supplement~B [\citet{frankssuppb}].}\label{figguardingsnapshots}
\end{figure}

As a sensitivity analysis, we fit EM in 100 different games, on
different teams, using only 30 possessions for estimating the
parameters of the model. The results show that thirty possessions are
enough to learn the weights to reasonable precision and that they are
stable across games: $\hat{\Gamma} = (0.62\pm0.02, 0.11\pm0.01,
0.27\pm0.02)$. {Values of the transition parameter are more
variable but have a smaller impact on inferred defensive matchups:
values range from $\rho= 0.96$ to $\rho= 0.99$.} Empirically, the
algorithm does a good job of capturing who's guarding whom. Figure~\ref
{figguardingsnapshots} illustrates a few snapshots from the
model. While there is often some uncertainty about who's guarding
whom near the basket, the model accurately infers switches and double
teams. See Supplement~B for animations demonstrating
the model performance [\citet{frankssuppb}].

%

This model is clearly interesting in its own right, but, most
importantly, it facilitates a plethora of new analyses which
incorporate matchup defense. For instance, the model could be
used to improve \emph{counterpart statistics}, a measure of how well
a player's counterpart performs [\citet{kubatko}]. Our
model circumvents the challenges associated with identifying the most
appropriate counterpart for a player, since we directly infer who is
guarding whom at every instant of a possession.

The model can also be used to identify how much defensive attention
each offender receives. Table~\ref{tabattention} shows the league
leaders in attention received, when possessing the ball and when not
possessing the ball. We calculate the average attention each player
receives as the total amount of time guarded by all defenders divided
by the total time playing. This metric reflects the perceived threat
of different offenders. The measure also provides a quantitative
summary of exactly how much a superstar may free up other shooters on
his team, by drawing attention away from them.

%
\begin{table}[b]
\caption{Average attention drawn, on and off ball. Using inference
about who's
guarding whom, we calculate the average attention each player
receives as the total amount of time guarded by each defender divided by
the total time playing (subset by time with and without the ball). At
any moment in time, there are five defenders, and hence five units of
``attention'' to divide among the five offenders
each possession. On ball, the players receiving the most attention
are double teamed an average of 20\% of their time possessing the
ball. Off ball, the players that command the most attention
consist largely of MVP caliber players
}
\label{tabattention}
\tabcolsep=0pt
\begin{tabular*}{\tablewidth}{@{\extracolsep{\fill}}@{}lcccc@{}}
\hline
& \multicolumn{2}{c}{\textbf{On ball}} & \multicolumn{2}{c@{}}{\textbf{Off ball}}\\[-6pt]
& \multicolumn{2}{c}{\hrulefill} & \multicolumn{2}{c@{}}{\hrulefill}\\
\textbf{Rank} & \textbf{Player} & \textbf{Attention}  &\textbf{Player} & \textbf{Attention}\\
\hline
1 & DeMar DeRozan & 1.213 &       Stephen Curry & 1.064 \\
2 & Kevin Durant & 1.209 &        Kevin Durant & 1.063 \\
3 & Rudy Gay & 1.201     &        Carmelo Anthony & 1.048 \\
4 & Eric Gordon & 1.187 &         Dwight Howard & 1.044 \\
5 & Joe Johnson & 1.181 &         Nikola Pekovic & 1.036 \\
\hline
\end{tabular*}
%
\end{table}

Alternatively, we can define some measure of \emph{defensive entropy}:
the uncertainty associated with whom a defender is guarding throughout
a possession. This may be a useful notion, since it reflects how
active a defender is on the court, in terms of switches and double
teams. If each defender guards only a single player throughout
the course of a possession, the defensive entropy is zero. If they
split their time equally between two offenders, their entropy is one.
Within a possession, we define a defender's entropy as $\sum_{k=1}^5
Z_n(j,k)\log(Z_n(j,k))$, where $Z_n(j,k)$ is the fraction of time
defender $j$
spends guarding offender $k$ in possession $n$.

By averaging defender entropy over all players on a defense, we get a
simple summary of a team's tendency for defensive switches and double teams.
Table~\ref{tabentropy} shows average team entropies, averaged over
all defenders within a defense as well as a separate measure averaging
over all defenders faced by an offense (induced entropy). By this
measure, the Miami Heat were the most active team defense, and,
additionally, they induce
the most defensive entropy as an offense.

%
\begin{table}[b]
\tabcolsep=0pt
\tablewidth=255pt
\caption{Team defensive entropy. A player's defensive entropy for a
particular possession is defined as $\sum_{k=1}^5
Z_n(j,k)\operatorname{log}(Z_n(j,k))$, where $Z_n(j,k)$
is the fraction of time the defender $j$ spends guarding offender $k$ during
possession $n$. Team defensive entropy is defined as the average
player entropy over all defensive possessions for that team. Induced
entropy is the average player
entropy over all defenders facing a particular offense}\label{tabentropy}
\begin{tabular*}{\tablewidth}{@{\extracolsep{\fill}}@{}lcc@{\qquad\qquad}lcc@{}}
\hline
& & & & & \textbf{Induced}\\
\textbf{Rank} & \textbf{Team} & \textbf{Entropy} & \textbf{Rank} & \textbf{Team} & \textbf{entropy}\\
\hline
\phantom{0}1 & Mia & 0.574    &        \phantom{0}1 & Mia & 0.535 \\
\phantom{0}2 & Phi & 0.568    &        \phantom{0}2 & Dal & 0.526 \\
\phantom{0}3 & Mil & 0.543    &        \phantom{0}3 & Was & 0.526 \\
\phantom{0}4 & Bkn & 0.538    &        \phantom{0}4 & Chi & 0.524 \\
\phantom{0}5 & Tor & 0.532    &        \phantom{0}5 & LAC & 0.522
\\[3pt]
26 & Cha & 0.433    &        26 & OKC & 0.440 \\
27 & Chi & 0.433    &        27 & NY & 0.440 \\
28 & Uta & 0.426    &        28 & Min & 0.431 \\
29 & SA & 0.398     &        29 & Phi & 0.428 \\
30 & Por & 0.395    &        30 & LAL & 0.418 \\
\hline
\end{tabular*}
\end{table}


{These results illustrate the many types of analyses that can be
conducted with this model, but there are still many ways in which
the model itself could be extended. By exploiting situational
knowledge of basketball, we could develop more complex and precise
models for the conditional defender behavior. In our model it is
theoretically simple to add additional covariates or latent
variables to the model which explain different aspects of team or
defender behavior. For instance, we could include a function of
defender velocity as an additional independent variable, with some
function of offender velocity as a covariate. Other covariates
might relate to more specific in game situations or only be
available to coaches who know the defensive game plan. Finally, by
including additional latent indicators, we could model defender
position as a mixture model over possible defensive schemes and
simultaneously infer whether a team is playing zone defense or man
defense. Since true zone defense is rare in the NBA, this approach
may be more appropriate for other leagues.}

{ We also make simplifying assumptions about homogeneity across
players. It is possible to account for heterogeneity across
players, groups of players, or teams by allowing the coefficients,
$\Gamma$, to vary in a hierarchy [see \citet{Maruotti2008} for a
related approach involving unit level random effects in HMM's].
Moreover, the hidden Markov model makes strong assumptions about the
amount of time each defender spends guarding a particular offender.
For instance, in basketball many defensive switches tend to be very
brief in duration, since they consist of quick ``help defense'' or a
short double team, before the defender returns to guarding their
primary matchup. As such, the geometric distribution of state
durations associated with the HMM may be too restrictive. Modeling
the defense with a hidden \emph{semi-Markov} model, which allows the
transition probabilities to vary as a function of the time spent in
each state, would be an interesting avenue for future research
[\citet{yu2010hidden,semi2001}].

While theoretically straightforward, these extensions require
significantly more computational resources. Not only are there more
coefficients to estimate, but as a consequence the algorithm must be
executed on a much larger set of possessions to get reasonable
estimates for these coefficients. Nevertheless, our method, which
ignores some of these complexities, passes the ``eye test''
(Figure~\ref{figguardingsnapshots}, Supplement~B [\citet{frankssuppb}])
and leads to
improved predictions about shot outcomes (Table~\ref{tabacc}).}

In this paper we emphasize the use of matchup
defense for inferring individual spatially referenced defender skill.
Using information about how long defenders guard offenders and who
they are guarding at the moment of the shot, we can estimate how
defenders affect both shot selection and shot efficiency in different
parts of the court. Still, given the high resolution of the spatial
data and relatively low sample size per player, inference is
challenging. As such, before proceeding we find an interpretable,
data-driven, low-dimensional spatial representation of the court on
which to estimate these defender effects.


%
%
\newcommand{\R} {\mathbb{R}}
\newcommand{\distNorm} {\mathcal{N}}

\section{Parameterizing shot types}\label{secnmf}

In order to concisely represent players' spatial offensive and
defensive ability, we develop a method to find a succinct
representation of the court by using the locations of attempted shots.
Shot selection in professional basketball is highly structured. We
leverage this structure by finding a low-dimensional decomposition of
the court whose components intuitively corresponds to \emph{shot
type}. {A \emph{shot type} is a cluster of ``similar'' shots
characterized by a spatially smooth intensity surface over the court.
This surface indicates where shots from that cluster tend to come from
(and where they do not come from)}. Each player's shooting habits are
then represented by a positive linear combination of the global shot types.

Defining a set of global shot types shared among players is beneficial
for multiple reasons. First, it allows us to concisely parameterize
spatial phenomena with respect to shot type (e.g., the ability
of a defensive player to contest a corner three-point shot). Second, it
provides a low-dimensional representation of player habits that can be
used to specify a prior on both offensive and defensive parameters for
possession outcomes. The graphical and numerical results of this model
can be found in Section~\ref{secnmf-results}.

\subsection{Point process decomposition}
Our goal is to simultaneously identify a small set of $\mathcal{B}$
global \emph{shot types} and each player's loadings onto these shot
types. We accomplish this with a two-step procedure. First, we find a
nonparametric estimate of each player's smooth intensity surface,
modeled as a log Gaussian Cox process (LGCP) [\citet{moller1998log}].
Second, we find an optimal low-rank representation of all players'
intensity surfaces using nonnegative matrix factorization (NMF)
[\citet
{lee1999learning}]. The LGCP incorporates individual spatial
information about shots, while NMF pools together global information
across players. This pooling smooths each player's estimated intensity
surface and yields more robust generalization. For instance, for
$\mathcal{B}=6$, the average predictive ability across players of
LGCP${}+{}$NMF outperforms the predictive ability of independent LGCP
surfaces on out-of-sample data. Intuitively, the global bases define
long-range correlations that are difficult to capture with a stationary
covariance function.

We model a player's shot attempts as a point process on the offensive
half court, a 47 ft by 50 ft rectangle. Again, shooters will be indexed
by~${k \in\{1, \dots, K\}}$, and the set of each player's shot
attempts will be referred to as~$\mathbf x_k = \{ x_{k,1}, \dots,
x_{k,N_k} \}$, where $N_k$ is the number of shots taken by player $k$,
and ${x_{k,m} \in[0,47]\times[0,50]}$.

Though we have formulated a continuous model for conceptual simplicity,
we discretize the court into~$V$ one-square-foot tiles for
computational tractability of LGCP inference. We expect this tile size
to capture all interesting spatial variation. Furthermore, the
discretization maps each player into $\R_{+}^V$, which is necessary
for the NMF dimensionality reduction.

Given point process realizations for each of $K$ players, $\mathbf x_1,
\dots, \mathbf x_K$, our procedure is
as follows:
\begin{enumerate}[2.]
\item Construct the count matrix $\mathbf X_{kv} ={}$number of shots by
player $k$ in tile $v$ on a discretized court.

%
%
%
\begin{figure}[b]
\begin{tabular}{@{}ccc@{}}

\includegraphics{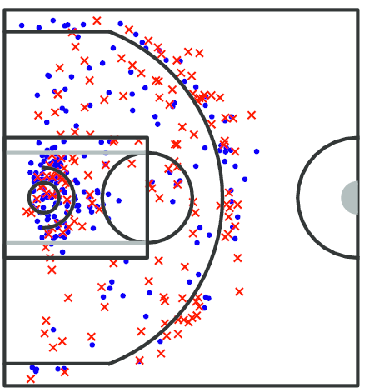}
 & \includegraphics{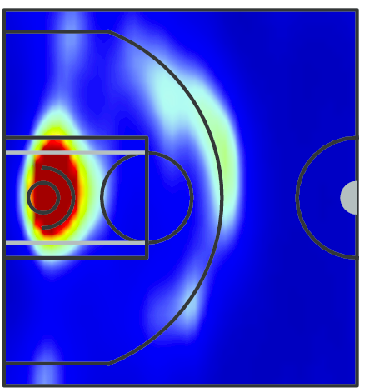} & \includegraphics{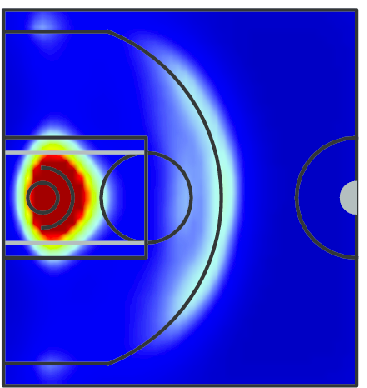}\\
\footnotesize{(a) Shots} & \footnotesize{(b) LGCP} & \footnotesize{(c) LGCP${}+{}$NMF}\\
\multicolumn{3}{@{}c@{}}{\footnotesize{LeBron James}}
\\[12pt]

\includegraphics{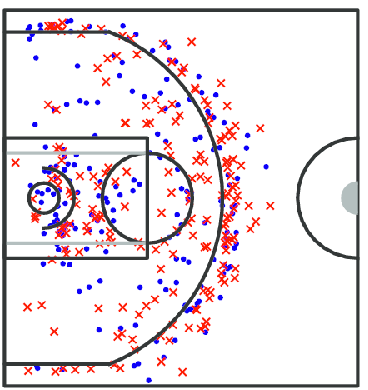}
 & \includegraphics{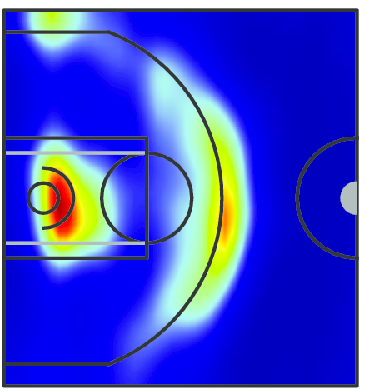} & \includegraphics{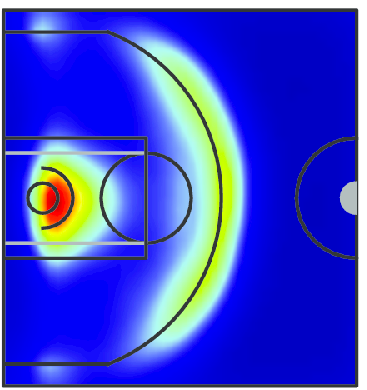}\\
\footnotesize{(d) Shots} & \footnotesize{(e) LGCP} & \footnotesize{(f) LGCP${}+{}$NMF}\\
\multicolumn{3}{@{}c@{}}{\footnotesize{Stephen Curry}}
\end{tabular}
\caption{NBA player shooting representations, from left to right:
original point process data from two players, LGCP surface, and NMF
reconstructed surfaces ($B=6$). Made and missed shots are represented
as blue circles and red $\times$'s, respectively.}\label{figlgcptonmf}\label{figLGCP}\label{figLGCPNMF}\label{figshots}
\end{figure}

\item Fit an intensity surface $\lambda_k = (\lambda_{k1}, \dots,
\lambda_{kV})^T$ for each player $k$ over the discretized court (LGCP)
[Figure~\ref{figLGCP}(b)].
\item Construct the data matrix $\bolds\Lambda= (\bar\lambda_1,
\dots, \bar\lambda_K)^T$, where $\bar\lambda_k$ has been
normalized to have unit volume.
\item Find low-rank matrices $\mathbf L, \mathbf W$ such that $\mathbf
W \mathbf L \approx\bolds\Lambda$, constraining all matrices to be
nonnegative (NMF) [Figure~\ref{figLGCPNMF}(c)].
\end{enumerate}

This procedure yields a spatial basis $\mathbf L$ and basis loadings,
$\hat{\mathbf w}_k$, for each individual player.

One useful property of the Poisson process is the superposition theorem
[e.g., \citet{kingman1992poisson}], which states that given a countable
collection of independent Poisson processes $\mathbf x_1, \mathbf x_2,
\dots,$ each with intensity $\lambda_1, \lambda_2, \dots,$ their
superposition, {defined as the union of all observations,} is
distributed as
\begin{eqnarray*}
\bigcup_{i=1}^\infty\mathbf x_i
&\sim& \mathcal{PP} \Biggl( \sum_{i=1}^\infty
\lambda_i \Biggr).
\end{eqnarray*}
%

Consequently, with the nonnegativity of the basis and loadings from
the NMF procedure, the basis vectors can be interpreted as
sub-intensity functions, or ``shot types,'' which are archetypal
intensities used by each player. The linear weights for each player\vspace*{1pt}
concisely summarize the spatial shooting habits of a player into a
vector in $\R_+^\mathcal{B}$.

\subsection{Fitting the LGCPs}
For each player's set of points, $\mathbf x_k$, the likelihood of the
point process is discretely approximated as
\begin{eqnarray*}
p\bigl(\mathbf x_k | \lambda_k(\cdot)\bigr) &\approx&
\prod_{v=1}^{V} p_{\mathrm{pois}}(\mathbf
X_{kv} | \Delta A \lambda_{kv} ),
\end{eqnarray*}
where, overloading notation, $\lambda_k(\cdot)$ is the exact
intensity function, $\lambda_k$ is the discretized intensity function
(vector), $\Delta A$ is the area of each tile (implicitly one from now
on), and $p_{\mathrm{pois}}(\cdot| \lambda)$ is the Poisson probability
{mass} function with mean $\lambda$. This approximation comes from the
completely spatially random property of the Poisson process, which
renders disjoint subsets of space independent. Formally, for two
disjoint subsets $A,B \subset\mathcal{X}$, after conditioning on the
intensity, the number of points that land in each set, $N_A$ and $N_B$,
are independent. Under the discretized approximation, the probability
of the number of shots in each tile is Poisson, with uniform intensity
$\lambda_{kv}$.

Explicitly representing the Gaussian random field $\mathbf z_k$, the
posterior is
\begin{eqnarray*}
p(\mathbf z_k | \mathbf x_k) &\propto& p(\mathbf
x_k | \mathbf z_k) p(\mathbf z_k)
\\
&=& \prod_{v=1}^{V} e^{-\lambda_{kv}}
\frac{\lambda_{kv}^{\mathbf
X_{kv}}}{\mathbf X_{kv}!} \distNorm( \mathbf z_k | 0, \mathbf C),
\\
\lambda_{n} &=& \exp( \mathbf z_k + z_0 ),
\end{eqnarray*}
where the prior over $\mathbf z_k$ is a mean zero normal with covariance
\[
\mathbf C_{vu} \equiv c(\mathbf{x}_v,
\mathbf{x}_u) = \sigma^2 \exp\Biggl( -\frac{1}{2}
\sum_{d=1}^2 \frac{(x_{vd} - x_{ud})^2}{\nu
_d^2} \Biggr)
\]
and $z_0$ is an intercept term that parameterizes the mean rate of the
Poisson process. This kernel is chosen to encode prior belief in the
spatial smoothness of player habits. Furthermore, we place a gamma
prior over the length scale, $\nu_k$, for each individual player. This
gamma prior places mass dispersed around 8 feet, indicating the
reasonable a priori belief that shooting variation is locally smooth on
that scale. Note that $\nu_k = (\nu_{k1}, \nu_{k2})$, corresponding
to the two dimensions of the court. We obtain posterior samples of
$\lambda_k$ and $\nu_k$ by iteratively sampling $\lambda_k | \mathbf
{x}_k, \nu_k$ and $\nu_k | \lambda_k, \mathbf{x}_k$.

We use Metropolis--Hastings to generate samples of $\nu_k | \lambda_k,
\mathbf{x}_k$. Details of the sampler are included in  Supplement~A
[\citet{frankssuppa}].

\subsection{NMF optimization}
Identifying nonnegative linear combinations of global shot types can
be directly mapped to nonnegative matrix factorization. NMF assumes
that some matrix $\bolds\Lambda$, in our case the matrix of
player-specific intensity functions, can be approximated by the product
of two low-rank matrices
\[
\bolds\Lambda= \mathbf W \mathbf L,
\]
where $\bolds\Lambda\in\R_+^{N \times V}$, $\mathbf{W} \in\R
_+^{N \times\mathcal{B}}$, and $\mathbf L \in\R_+^{\mathcal{B}
\times V}$, and we assume $\mathcal{B} \ll V$. The optimal matrices
$\mathbf W^*$ and $\mathbf L^*$ are determined by an optimization
procedure that minimizes $\ell(\cdot, \cdot)$, a measure of
reconstruction error or divergence between $\mathbf W \mathbf L$ and
$\bolds\Lambda$ with the constraint that all elements remain nonnegative,
\begin{eqnarray*}
\mathbf W^*, \bolds\ell^* &=& \mathop{\arg\min}_{\mathbf W_{ij}, \mathbf
L_{ij} \geq0} \ell(\bolds
\Lambda, \mathbf W \mathbf L).
\end{eqnarray*}
Different choices of $\ell$ will result in different matrix
factorizations. 
A natural choice is the matrix divergence metric
\begin{eqnarray*}
\ell_{\mathrm{KL}}(\mathbf A, \mathbf B) &=& \sum_{i,j}
X_{ij} \log\frac{A_{ij}}{B_{ij}} - A_{ij} + A_{ij},
\label{eqkl}
\end{eqnarray*}
{which corresponds to the Kullback--Leibler (KL) divergence if $\mathbf
A$ and $\mathbf B$ are discrete distributions, that is, $\sum_{ij}
A_{ij} = \sum_{ij} B_{ij} = 1$ [\citet{lee}}]. Although there are
several other possible divergence metrics (i.e., Frobenius), we use
this KL-based divergence measure for reasons outlined in \citet
{miller2014}. 
We solve the optimization problem using techniques from \citet
{lee} and \citet{brunet}.

Due to the positivity constraint, the basis $\mathbf L^*$ tends to be
disjoint, exhibiting a more ``parts-based'' decomposition than other,
less constrained matrix factorization methods, such as PCA. This is due
to the restrictive property of the NMF decomposition that disallows
negative bases to cancel out positive bases. In practice, this
restriction eliminates a large swath of ``optimal'' factorizations with
negative basis/weight pairs, leaving a sparser and often more
interpretable basis [\citet{lee1999learning}].


\subsection{Basis and player summaries}
\label{secnmf-results}
%
%
%
\begin{figure}[b]

\includegraphics{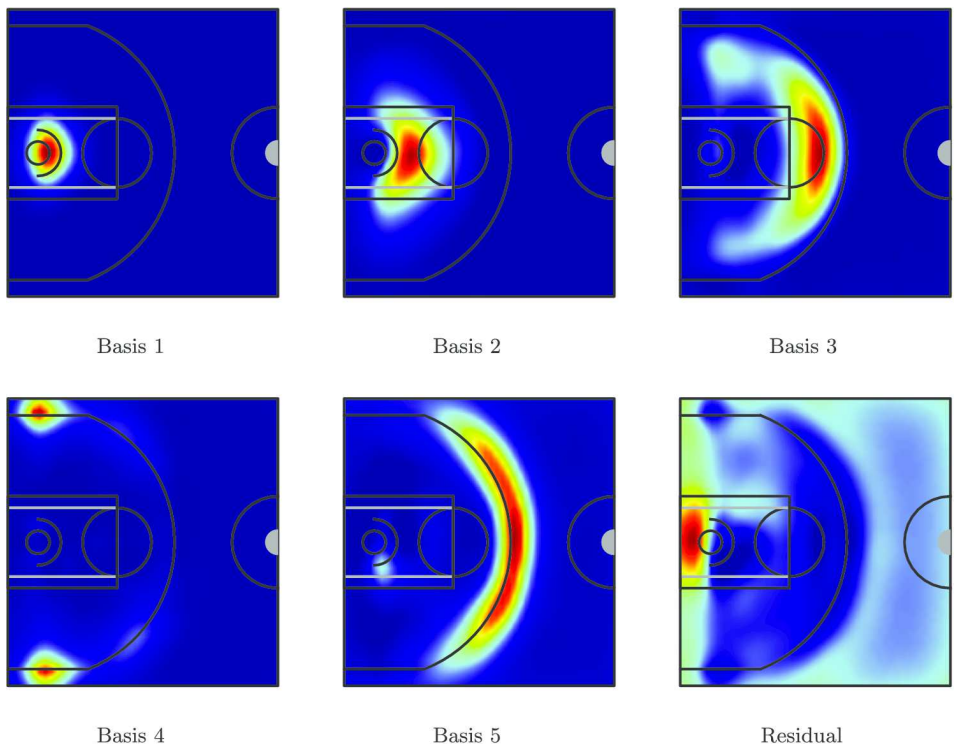}

\caption{Basis vectors (surfaces) identified by LGCP--NMF for $\mathcal
{B}=6$. Each basis surface is the normalized intensity function of a
particular shot type, and players' shooting habits are a weighted
combination of these shot types. Conditioned on a certain shot type
(e.g., corner three), the intensity function acts as a density over
shot locations, where red indicates likely locations.}\label{figbasis}
\end{figure}


We graphically depict the shot type preprocessing procedure in
Figure~\ref{figlgcptonmf}. A player's spatial shooting habits are
reduced from a raw point process to an independent intensity surface,
and finally to a linear combination of $\mathcal{B}$ nonnegative
basis surfaces. There is wide variation in shot selection among NBA
players---some shooters specialize in certain types of shots, whereas
others will shoot from many locations on the court.

{We set ${\mathcal{B}=6}$ and use the KL-based loss function,
choices which exhibit sufficient predictive ability in \citet
{miller2014}, and yield an interpretable basis. We graphically depict
the resulting basis vectors in Figure~\ref{figbasis}.} This procedure
identifies basis vectors that correspond to spatially interpretable
shot types. Similar to the parts-based decomposition of human faces
that NMF yields in \citet{lee1999learning}, LGCP--NMF yields a
shots-based decomposition of NBA players. For instance, it is clear
from inspection that one basis corresponds to shots in the restricted
area, while another corresponds to shots from the rest of the paint.
The three-point line is also split into corner three-point shots and
center three-point shots. Unlike PCA, NMF is not mean centered, and, as
such, a residual basis appears regardless of $\mathcal{B}$; this basis
in effect captures positive intensities outside of the support of the
relevant bases. In all analyses herein, we discard the residual basis
and work solely with the remaining bases.

The LGCP--NMF decomposition also yields player-specific shot weights
that provide a concise characterization of their offensive habits. The
weight $w_{kb}$ can be interpreted as the amount player $k$ takes shot
type $b$, which quantifies intuitions about player behavior. These
weights will be incorporated into an informative prior over offensive
skill parameters in the possession outcome model.
{We highlight individual player breakdowns in Supplement~A [\citet
{frankssuppa}].} While these weights summarize offensive \emph{habits},
our aim is to develop a model to jointly measure both offensive and
defensive \emph{ability} in different parts of the court. Using who's
guarding whom and this data-driven court discretization, we proceed by
developing a model to quantify the effect that defenders have on both
shot selection (frequency) and shot efficiency.


%
%

\section{Frequency and efficiency: Characteristics of a shooter}

We proceed by decomposing a player's habits in terms of shot
frequency and efficiency. First, we construct a model for where on
the court different offenders prefer to shoot. This notion is often
portrayed graphically as the \textit{shot chart} and reflects a player's
spatial shot \textit{frequency}. Second, conditioned on a player
taking a shot, we want to know the probability that the player
actually makes the shot: the spatial player \textit{efficiency}.
Together, player spatial shot \emph{frequency} and \emph{efficiency} largely
characterize a basketball player's habits and ability.

While it is not difficult to empirically characterize frequency and
efficiency of shooters, it is much harder to say something about how
defenders affect these two characteristics. Given knowledge of
matchup defense, however, we can create a more sophisticated
joint model which incorporates how defenders affect shooter
characteristics. Using the results on who's guarding whom, we are
able to provide estimates of defensive impact on shot frequency and
efficiency, and ultimately a defensive analogue to the offensive shot
chart [Figure~\ref{figshots}(a)].

\subsection{Shrinkage and parameter regularization}
\label{secshrink}
Parameter regularization is a very important part of our model because
many players are only observed in a handful of plays. We shrink
estimates by exploiting the notion that players with similar roles
should be more similar in their capabilities. However, because
offense and defense are inherently different, we must characterize
player similarity separately for offense and defense.

%
\begin{figure}

\includegraphics{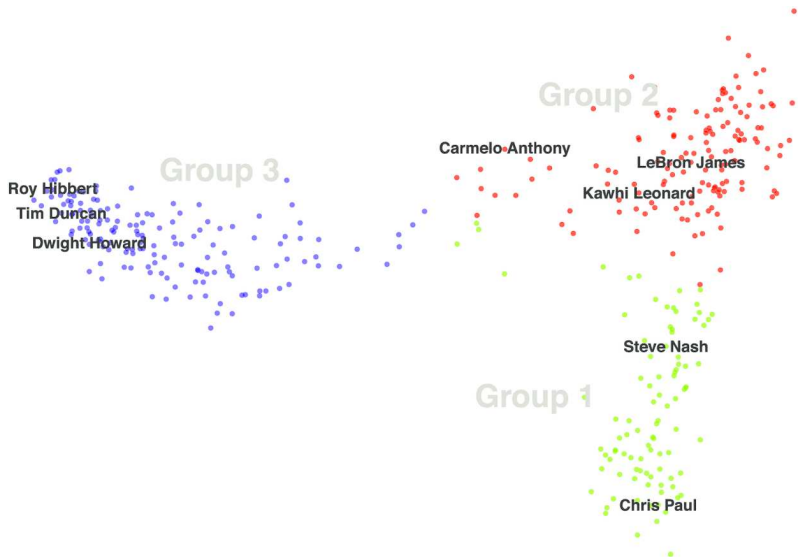}

\caption{Defensive slusters. We ran SVD on the $N \times\mathcal{B}$
matrix of time spent in each basis. The \mbox{$x$- }and $y$-axis correspond to
principal components one and two of this matrix. The first two principal
components suggest that three clusters reasonably separate player
groups. Group 1 (green) roughly corresponds to small point guards,
group 2 (red) to forwards and guards, and group 3 (blue) to centers.}\label{dclusts}
\end{figure}

First, we gauge how much variability there is between defender types.
One measure of defender characteristics is the fraction of time, on
average, that each defender spends guarding a shooter in each of the
$\mathcal{B}$ bases. Figure~\ref{dclusts} suggests that defenders can
be grouped into roughly three defender types. The groupings are
inferred using three cluster K-means on the first two principal
component vectors of the ``time spent'' matrix. Empirically, group 1
corresponds to small point guards, group 2 to forwards and guards, and
group 3 to centers. We use these three groups to define the shrinkage
points for defender effects in both the shot selection and shot
efficiency models.

When we repeat the same process for offense, it is clear that the
players do not cluster; specifically, there appears to be far more
variability in offender types than defender types. Thus, to
characterize offender similarity, we instead use the normalized player
weights from the nonnegative matrix factorization, $\mathbf{W}$, introduced
in Section~\ref{secnmf} and described further in  Supplement~A
[\citet{frankssuppa}].
Figure~\ref{oclusts} shows the loadings on the
first two principal components of the player weights. The points are
colored by the player's listed position (e.g., guard, center, forward,
etc.). While players tend to be more similar to players with
the same listed position, on the whole, position is not a good
predictor of an offender's shooting characteristics.

%
\begin{figure}

\includegraphics{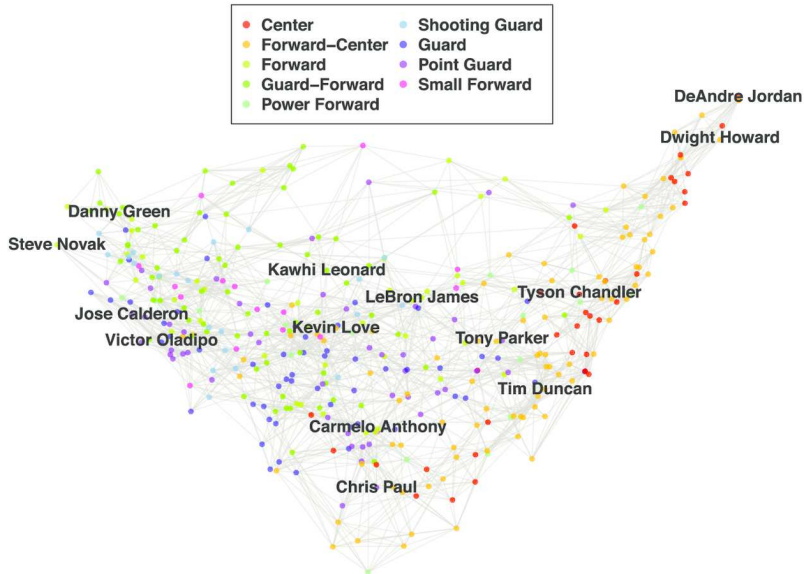}

\caption{Offender similarity network. We ran SVD on the $N \times
\mathcal{B}$ matrix of NMF coefficients (Section~\protect\ref{secnmf}).
The $x$- and $y$-axis correspond to principal components one and two of
this matrix. The projection into the first two principal components
shows that there is no obvious clustering of offensive player types,
as was the case with defense. Moreover, ``player position'' is not a
good indicator of shot selection.}
\label{oclusts}
\end{figure}

Consequently, for the prior distribution on offender efficiency we use
a normal conditional autoregressive (CAR) model [\citet{cressie93}].
For every player, we identify the 10 nearest neighbors in the space of
shot selection weights. We then connect two players if, for either
player in the pair, their partner is one of their ten closest
neighbors. We use this network to define a Gaussian Markov random
field prior on offender efficiency effects (Section~\ref{secefficiency}).




\subsection{Shot frequency}

We model shot selection (both shooter and location) using a
multinomial distribution with a logit link function. First, we
discretize the court into $\mathcal{B}$ regions using the
preprocessed NMF basis vectors (see Section~\ref{secnmf}) and define
the multinomial outcomes as one of the $5\times\mathcal{B}$
shooter/basis pairs. The court regions from the NMF are naturally
disjoint (or nearly so). In this paper, we use the first five bases
given in Figure~\ref{figbasis}. Shot selection is a function of the
offensive players on the court, the fraction of possession time that
they are guarded by different defenders, and defenders' skills.
Letting $\mathcal{S}_n$ be a categorical random variable indicating
the shooter and shot location in possession $n$,
\[
p\bigl( \mathcal{S}_n(k,b)=1 | \alpha, Z_{n}\bigr) =
\frac{\exp( \alpha_{kb} + \sum_{j=1}^5 Z_n(j,k)\beta
_{jb} )}{1 + \sum_{mb} \exp( \alpha_{kb} + \sum_{j=1}^5
Z_n(j,k) \beta_{jb}) }.
\]

Here, $\alpha_{kb}$ is the propensity for an offensive player, $k$, to
take a shot from basis~$b$. However, in any given possession, a
players' propensity to shoot is affected by the defense. $\beta_{jb}$
represents how well a defender, $j$, suppresses shots in a
given basis $b$, relative to the average defender in that basis. These values
are modulated by entries in a possession specific covariate matrix
$Z_n$. The value $Z_n(j,k)$ is the fraction of time defender
$j$ is guarding offensive player $k$ in possession $n$, with $\sum_{k=1}^5
Z_n(j,k) = 1$. We infer $Z_n(j,k)$ for each possession using the
defender model outlined
in Section~\ref{secguarding}. Note that the baseline outcome is ``no
shot,'' indicating there was a turnover before a shot was attempted.

We assume normal random effects for both the offensive and
defensive player parameters:
\[
\alpha_{kb} \sim N\bigl(\mu_{\alpha b},\sigma^2_{\alpha}
\bigr),\qquad\beta_{jb} \sim N\bigl(\mu_{\beta\mathcal{G}b},
\sigma^2_{\beta}\bigr).
\]
Here, $\mu_{\alpha b}$ and $\mu_{\beta\mathcal{G}b}$ represent
the player average effect in basis $b$ on offense and defense,
respectively. For defenders, $\mathcal{G}$ indexes one of the 3
defender types (Figure~\ref{dclusts}), so that there are in fact
3$\mathcal{B}$ group
means. Finally, we specify that
\[
\mu_{\alpha b} \sim N\bigl(0, \tau^2_{\alpha}\bigr),
\qquad\mu_{\beta\mathcal{G}b} \sim N\bigl(0, \tau^2_{\beta}
\bigr).
\]
%

\subsection{Shot efficiency}\label{secefficiency}

Given a shot, we model efficiency (the probability that the shot is
made) as a function of the offensive player's skill, the defender at
the time of the shot, the distance of that defender to the shooter,
and where the shot was taken. For a possession $n$,
\[
p\bigl( Y_n=1 | \mathcal{S}_n(k,b)=1, j,
\mathcal{D}_n, \theta,\phi,\xi\bigr) = \frac{\operatorname{exp}( \theta
_{kb} +
\phi_{jb}+\xi_{b}\mathcal{D}_n )}{1+\operatorname{exp}( \theta
_{kb} +
\phi_{jb}+\xi_{b}\mathcal{D}_n)}.
\]

\begin{figure}

\includegraphics{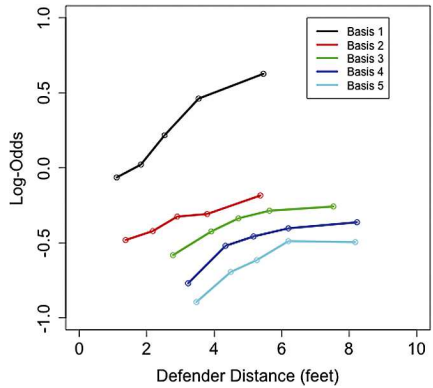}

\caption{Shot efficiency vs. distance. We plot empirical shot
efficiency as a function of the guarding defender's distance, by
region. 
We compute the
empirical log-odds of a shot by binning all shots from each region
into 5 bins. {Within region,} between 0 and 6~ft the log-odds
of a made shot appears to be nearly linear in distance. {After
about 6~ft (depending on the basis), increased defender distance
does not continue to increase the odds of a made shot.}}
\label{defender-dist}
\end{figure}

Here, $Y_n$ is an indicator for whether the attempted shot for
possession $n$ was made and $\mathcal{D}_n$ is the distance in feet
between the shooter and defender at the moment of the shot, capped at
some inferred maximum distance. The parameter $\theta_{kb}$ describes
the shooting skill of a player, $k$, from basis $b$. {The two
terms, $\phi_{jb}$ and $\xi_{b}\mathcal{D}_n$, are meant to represent
orthogonal components of defender skill. $\phi_{jb}$ encompasses how
well the defender contests a shot regardless of distance,
$\xi_{b}\mathcal{D}_n$ is independent of the defender identity and
adjusts for how far the defender is from the shot.} Within a
region, as the defender gets farther from the shooter, their effect on
the outcome of the shot decreases at the same rate, $\xi_{b}$; as the
most likely defender approaches the exact location of the shooter, the
defensive effect on the log-odds of a made shot converges toward
$\phi_{jb}$. Figure~\ref{defender-dist} supports this modeling
choice: empirically, the log-odds of a shot increase roughly linearly
in distance up until a point (around 5 or 6 feet depending on the
region) at which distance no longer has an
effect.

%

We again employ hierarchical priors to pool information
across players. On defense we specify that
\[
\phi_{jb} \sim N\bigl(\mu_{\phi\mathcal{G}b},\sigma
^2_{\phi}\bigr).
\]
Here, $\mu_{\phi\mathcal{G}b}$ represents the player
average effect in basis $b$ on defense.
Again, $\mathcal{G}$ indexes one of 3 defender types, so that
there are in fact 3$\mathcal{B}$ group means.

On offense, we use the network defined in Section~\ref{secshrink}
(Figure~\ref{oclusts}) to specify a
CAR prior. We define each player's efficiency to be, a
priori, normally distributed with mean proportional to the mean of his
neighbors' efficiencies. This operationalizes the notion that
players who have more similar shooting habits should have more similar
shot efficiencies. Explicitly, the efficiency $\theta$ of an offender
$k$ in a region $b$ with mean player efficiency
$\mu_{\theta b}$ has the prior distribution
\[
(\theta_{kb}-\mu_{\theta b}) \sim N \biggl(\frac{\zeta}{|\mathcal
{N}(k)|}\sum
_{k' \in
\mathcal{N}(k)}(\theta_{k'b}-\mu_{\theta b}),
\sigma_{k}^2 \biggr),
\]
where $\mathcal{N}(k)$ are the set of neighbors for offender $k$ and
$\zeta\in[0,1)$ is a discount factor. These conditionals imply the
joint distribution
\[
\bolds{\theta}_b \sim N\bigl(\bolds{\mu_{\theta b}},(I-\zeta
M)^{-1}D\bigr),
\]
where\vspace*{-2pt} $D$ is the diagonal matrix with
entries $\frac{1}{\sigma^2_{k}}$ and $M$ is the matrix such that
$M_{k,k'} = \frac{1}{|\mathcal{N}(k)|}$ if offenders $k$ and $k'$ are
neighbors and zero otherwise. This joint distribution is proper as
long as $(I-\zeta M)^{-1}D$ is symmetric positive-definite. The
matrix is symmetric when $\sigma^2_{k} \propto
\frac{1}{\mathcal{N}(k)}$. We\vspace*{2pt} chose $\zeta=0.9$ to guarantee the
matrix is positive-definite [\citet{cressie93}]. { The number of
neighbors (Figure~\ref{oclusts}) determines the shrinkage point for
each player and $\zeta$ control how much shrinkage we do. We chose
the number of neighbors to be relatively small and hence the $\zeta$
to be relatively large, since the players in a neighborhood should
be quite similar in their habits. }

Again we use normal priors for the group means:
\[
\mu_{\theta b} \sim N\bigl(0, \tau^2_{\theta}\bigr),\qquad
\mu_{\phi b} \sim N\bigl(0, \tau^2_{\phi}\bigr).
\]

Finally, for the distance effect, we specify that
\[
\xi_b \sim N^+\bigl(0,\tau^2_\xi\bigr),
\]
where $N^+$ indicates a half-normal distribution. We chose a prior
distribution with positive support, since increased defender distance
should logically increase the offenders' efficiency.

\subsection{Inference}

We use Bayesian inference to infer parameters of both the shot
frequency and shot efficiency models. First, we consider different
methods of inference in the shot frequency model. The sample size,
number of categories and number of parameters in the model for shot
selection are all quite large, making full Bayesian inference
challenging. Specifically, there are $5 \times\mathcal{B} + 1 = 26$
outcomes (one for each
shooter-basis pair plus one for turnovers) and nearly 150,000
observations. To facilitate computation, we use a local variational
inference strategy to approximate the true posterior of parameters
from the multinomial logistic regression. The idea behind the
variational strategy is to find a lower bound to the multinomial
likelihood with a function that looks Gaussian. For notational
simplicity let $\bolds{\eta}_{n}$ be the vector with elements
$\eta_{nk} = \alpha_{kb} + \sum_{j=1}^5 Z_n(j,k)\beta_{jb}$. Then,
the lower bound takes the form
\[
\operatorname{log}P(\mathcal{S}_n |\bolds{\eta}_n)
\geq(\mathcal{S}_n+b_n)^T\bolds{
\eta}_{n}-\bolds{\eta}_{n}^T\mathbf{A}\bolds{
\eta}_{n} - c_n,
\]
where $\mathbf{b_n}$ and $c_n$ are variational parameters and $\mathbf{A}$
is a simple bound on the Hessian of the log-sum-exp function
[\citet{bohning92}]. This implies a
Gaussianized approximation to the observation model. Since we use
normal priors on the parameters, this yields a normal approximation to
the posterior. By iteratively updating the variational parameters, we
maximize the lower bound on the likelihood. This yields the best
normal approximation to the posterior in terms of KL-divergence [see
\citet{murphyml} for details].

In the variational inference, we fix the prior parameters as follows:
$\sigma_\alpha^2 = 1$, $\sigma_\beta^2 = 0.01$, $\tau_\alpha^2 = 1$,
and $\tau_\beta^2 = 0.01$. That is,\vspace*{1pt} we specify more prior variability
in the offensive effects than the defensive effects at both the group
and individual level. We use cross-validation to select these prior
parameters, and then demonstrate that despite using approximate
inference, the model performs well in out-of-sample prediction
(Section~\ref{secresults}). Since the variational method is only
approximate, we start with some exploratory analysis to tune the
shrinkage hyperparameters. We examine five scales for both the
offense and defense group level prior variance to find the shrinkage
factors that yield the highest predictive power.\vspace*{2pt} Because the random
effects are normal and additive, we constrain $\sigma_\beta^2 <
\sigma_\alpha^2$ for identifiability. We then fix the sum
$\sigma_{\mathrm{total}} = \sigma_\alpha^2 + \sigma_\beta^2$ and search over
values such that $\sigma_\beta^2 < \sigma_\alpha^2$. We also examine
different scales of $\sigma_{\mathrm{total}}$. This search at multiple values
of $\sigma_{\mathrm{total}}$ yields the optimal ratio
$\frac{\sigma_\beta^2}{\sigma_\alpha^2}$ to be between $0.1$ and
$0.2$.

For the efficiency model, we found Bayesian logistic regression to be
more tractable: in this regression, there are only two outcomes (make or
miss) and approximately 115,000 possessions which lead to a shot. Thus,
we proceed
with a fully Bayesian regression on shot efficiency, using the
variational inference algorithm to initialization of the sampler.
Inference in the Bayesian regression for shot efficiency was done
using hybrid Monte Carlo (HMC) sampling. We implemented the sampler
using the probabilistic programming language STAN [\citet{stan}].
We use
2000 samples, and ensure that the $\hat{R}$ statistic is close to 1 for
all parameters [\citet{gelman92}].

%
%

\section{Results}
\label{secresults}

We fit our model on data from the 2013--2014 NBA
regular season, focusing on a specific subset of play: possessions
lasting at least 5 seconds, in which all players are in the
half-court. We also ignore any activity after the first shot and
exclude all plays including fouls or stoppages for simplicity.

%
\begin{table}
\tabcolsep=0pt
\caption{Out-of-sample log-likelihoods for models of increasing
complexity. The first row corresponds to the average out-of-sample
likelihood for predicting only the shooter. The second row similarly
summarizes out-of-sample likelihood for predicting only which basis the
shot comes from (not the shooter). The third row is the average
out-of-sample log-likelihood over the product space of shooter and shot
location. We demonstrate that our model not only outperforms simpler
models in predicting possession outcomes, but also outperforms them in
both shooter and basis prediction tasks individually. In the fourth
row, we display the out-of-sample likelihoods for shot efficiency
(whether the shooter makes the basket). The four different models from
left to right are \textup{(i)} the full offensive and defensive model with
parameter shrinkage (incorporating inferred defender type and offender
similarity), \textup{(ii)} the offensive and defensive model with a common
shrinkage point for all players, \textup{(iii)} the offense only model, \textup{(iv)} the
offense only model with no spatial component. Incorporating defensive
information, spatial information and player type clearly yields the
best predictive models. All quantities were computed using 10-fold
cross-validation}\label{tabacc}
\begin{tabular*}{\tablewidth}{@{\extracolsep{\fill}}@{}lcccc@{}}
\hline
& \textbf{Full model} & \textbf{No shrinkage} & \textbf{No defense} & \textbf{No spatial}\\
\hline
Shooter log-likelihood & $\bolds{-25{,}474.93}$ & $-$25{,}571.41 & $-$25{,}725.17 & $-$26{,}342.83 \\
Basis log-likelihood & $\bolds{-25{,}682.16}$ & $-$25{,}740.27 & $-$25{,}809.14 & N/A \\
Full log-likelihood & $\bolds{-41{,}461.74}$ & $-$41{,}646.81 & $-$41{,}904.48 & N/A
\\[3pt]
Efficiency log-likelihood & $\phantom{0,}\bolds{-3202.09}$ & \phantom{0,}$-$3221.44 & \phantom{0,}$-$3239.12 & \phantom{0,}$-$3270.99\\
\hline
\end{tabular*}\vspace*{5pt}
\end{table}

First, we assess the predictive performance of our model relative to
simpler models. For both the frequency and efficiency models, we run
10-fold cross-validation and compare four models of varying complexity:
(i) the full offense/defense model with defender types and CAR
shrinkage, (ii) the full \mbox{offense/defense} model without defender types
or CAR shrinkage, (iii) a model that ignores defense completely, (iv) a
model that ignores defense and space. The frequency models (i)--(iii) all
include 5 ``shot-types,'' and each possession results in one of 26
outcomes. Frequency model (iv) has only 6 outcomes---who shot the ball
(or no shot). The outcomes of the efficiency model are always binary
(corresponding to made or missed shots).

Table~\ref{tabacc} demonstrates that we outperform simpler models in
predicting out-of-sample shooter-basis outcomes. Moreover, while we do
well in joint prediction, we also outperform simpler models for
predicting both shooter and shot basis separately. Finally, we show
that the full efficiency model also improves upon simpler models.
Consequently, by incorporating spatial variation and defensive
information we have created a model that paints a more detailed and
accurate picture of the game of
basketball.

%
\begin{table}[t]
\caption{Basis 1. Shot efficiency (top table) and frequency (bottom
table). We list the top and bottom five defenders in terms of
the effect on the log-odds on a shooters' shot efficiency in the
restricted area (basis~1). Negative effects imply that the defender \emph{decreases} the
log-odds of an outcome, relative to the global average player (zero
effect). The three columns consist of defenders in the three groups listed
in Figure~\protect\ref{dclusts} and the respective group means. {Roy
Hibbert, considered one of the best defenders near the basket,
reduces shot efficiency there more than any other player. Chris
Paul, a league leader in steals, reduces opponents' shot frequency
more than any other player of his type}}\label{tabbasis1}
\tabcolsep=0pt
\begin{tabular*}{\tablewidth}{@{\extracolsep{\fill}}lc@{\qquad}lc@{\qquad}lc@{}}
\hline
\multicolumn{2}{@{}c}{\textbf{Group 1}} & \multicolumn{2}{c}{\textbf{Group 2}} & \multicolumn{2}{c@{}}{\textbf{Group 3}}\\[-6pt]
\multicolumn{2}{@{}c@{\qquad}}{\hrulefill} & \multicolumn{2}{c@{\qquad}}{\hrulefill} & \multicolumn{2}{c@{}}{\hrulefill}\\
\textbf{Player} & $\bolds{\phi+\xi\mathcal{D}^*}$ & \textbf{Player} & $\bolds{\phi+\xi\mathcal{D}^*}$ & \textbf{Player} & $\bolds{\phi+\xi\mathcal{D}^*}$\\
\hline
\multicolumn{6}{@{}c@{}}{\textit{Basis} 1---\textit{efficiency}}\\
J. Smith & $-$0.116 & Kidd--Gilchrist & $-$0.068 & R. Hibbert & $-$0.618 \\
J. Lin & $-$0.029 & K. Singler & \phantom{$-$}0.016 & E. Brand & $-$0.484 \\
K. Thompson & $-$0.011 & T. Evans & \phantom{$-$}0.017 & R. Lopez & $-$0.462 \\
P. Pierce & \phantom{$-$}0.024 & Antetokounmpo & \phantom{$-$}0.035 & A. Horford & $-$0.461 \\
E. Bledsoe & \phantom{$-$}0.034 & A. Tolliver & \phantom{$-$}0.040 & K. Koufos & $-$0.450
\\[3pt]
\emph{Average} & \phantom{$-$}0.191 & \emph{Average} & \phantom{$-$}0.142 & \emph{Average} & $-$0.170
\\[3pt]
B. Jennings & \phantom{$-$}0.358 & J. Meeks & \phantom{$-$}0.327 & C. Boozer & $-$0.017 \\
R. Rubio & \phantom{$-$}0.406 & J. Salmons & \phantom{$-$}0.334 & J. Adrien & \phantom{$-$}0.006 \\
J. Wall & \phantom{$-$}0.414 & C. Parsons & \phantom{$-$}0.344 & D. Cunningham & \phantom{$-$}0.045 \\
B. Knight & \phantom{$-$}0.452 & J. Harden & \phantom{$-$}0.375 & O. Casspi & \phantom{$-$}0.102 \\
J. Teague & \phantom{$-$}0.512 & E. Gordon & \phantom{$-$}0.524 & T. Young & \phantom{$-$}0.126
\\[18pt]
\hline
\multicolumn{2}{@{}c}{\textbf{Group 1}} & \multicolumn{2}{c}{\textbf{Group 2}} & \multicolumn{2}{c@{}}{\textbf{Group 3}}\\[-6pt]
\multicolumn{2}{@{}c@{\qquad}}{\hrulefill} & \multicolumn{2}{c@{\qquad}}{\hrulefill} & \multicolumn{2}{c@{}}{\hrulefill}\\
\textbf{Player} & $\bolds{\beta}$ & \textbf{Player} & $\bolds{\beta}$ & \textbf{Player} & $\bolds{\beta}$\\
\hline
\multicolumn{6}{@{}c@{}}{\textit{Basis} 1---\textit{frequency}}\\
C. Paul & $-$0.422 & L. Deng & $-$0.481 & L. Aldridge & $-$0.050 \\
G. Hill & $-$0.375 & L. Stephenson & $-$0.464 & C. Boozer & $-$0.039 \\
I. Thomas & $-$0.367 & A. Afflalo & $-$0.450 & N. Pekovic & $-$0.027 \\
C. Anthony & $-$0.344 & L. James & $-$0.449 & T. Thompson & $-$0.026 \\
K. Hinrich & $-$0.334 & H. Barnes & $-$0.432 & D. Lee & \phantom{$-$}0.005
\\[3pt]
\emph{Average} & $-$0.255 & \emph{Average} & $-$0.333 & \emph{Average} & \phantom{$-$}0.157
\\[3pt]
S. Marion & $-$0.144 & J. Dudley & $-$0.226 & A. Drummond & \phantom{$-$}0.313 \\
G. Dragic & $-$0.136 & P. George & $-$0.213 & S. Hawes & \phantom{$-$}0.327 \\
D. Lillard & $-$0.134 & A. Aminu & $-$0.191 & J. Henson & \phantom{$-$}0.338 \\
J. Smith & $-$0.133 & T. Ross & $-$0.186 & E. Kanter & \phantom{$-$}0.376 \\
B. Jennings & $-$0.132 & J. Meeks & $-$0.148 & R. Lopez & \phantom{$-$}0.470\\
\hline
\end{tabular*}\vspace*{3pt}
\end{table}

%
\begin{table}[t]
\caption{Basis 5. Shot efficiency (top table) and frequency (bottom
table). We list the top and bottom five defenders in terms of
the effect on the log-odds on a shooters' shot efficiency from
center three (basis 5). Negative effects imply that the defender \emph
{decreases} the
log-odds of an outcome, relative to the global average player (zero
effect). The
three columns consist of defenders in the three groups listed
in Figure~\protect\ref{dclusts} and the respective group means.
{Hibbert, who is the best
defender near the basket (Table~\protect\ref{tabbasis1}), is the
worst at
defending on the perimeter. His opponents have higher log-odds of
making a three-point shot against him, likely because he is late
getting out to the perimeter to contest shots}}\label{tabbasis5}
\tabcolsep=0pt
\begin{tabular*}{\tablewidth}{@{\extracolsep{\fill}}lc@{\qquad}lc@{\qquad}lc@{}}
\hline
\multicolumn{2}{@{}c}{\textbf{Group 1}} & \multicolumn{2}{c}{\textbf{Group 2}} & \multicolumn{2}{c@{}}{\textbf{Group 3}}\\[-6pt]
\multicolumn{2}{@{}c@{\qquad}}{\hrulefill} & \multicolumn{2}{c@{\qquad}}{\hrulefill} & \multicolumn{2}{c@{}}{\hrulefill}\\
\textbf{Player} & $\bolds{\phi+\xi\mathcal{D}^*}$ & \textbf{Player} & $\bolds{\phi+\xi\mathcal{D}^*}$ & \textbf{Player} & $\bolds{\phi+\xi\mathcal{D}^*}$\\
\hline
\multicolumn{6}{@{}c@{}}{\textit{Basis} 5---\textit{efficiency}}\\
D. Collison & $-$0.183 & C. Lee & $-$0.165 & B. Bass & $-$0.075 \\
S. Curry & $-$0.170 & D. Wade & $-$0.142 & D. Green & $-$0.060 \\
N. Cole & $-$0.165 & D. DeRozan & $-$0.137 & D. West & $-$0.032 \\
A. Bradley & $-$0.164 & J. Crawford & $-$0.117 & T. Jones & $-$0.016 \\
P. Mills & $-$0.149 & L. Stephenson & $-$0.114 & B. Griffin & \phantom{0}0.012
\\[3pt]
\emph{Average} & $-$0.055 & \emph{Average} & $-$0.030 & \emph{Average} & \phantom{0}0.073\\[3pt]
J. Holiday & \phantom{0}0.014 & J. Green & \phantom{0}0.053 & P. Millsap & \phantom{0}0.088 \\
J. Jack & \phantom{0}0.020 & C. Parsons & \phantom{0}0.055 & T. Gibson & \phantom{0}0.105 \\
D. Williams & \phantom{0}0.027 & M. Harkless & \phantom{0}0.060 & T. Thompson & \phantom{0}0.114 \\
J. Smith & \phantom{0}0.042 & J. Smith & \phantom{0}0.063 & A. Davis & \phantom{0}0.148 \\
M. Dellavedova & \phantom{0}0.062 & G. Hayward & \phantom{0}0.072 & L. Aldridge & \phantom{0}0.188
\\[18pt]
\hline
\multicolumn{2}{@{}c}{\textbf{Group 1}} & \multicolumn{2}{c}{\textbf{Group 2}} & \multicolumn{2}{c@{}}{\textbf{Group 3}}\\[-6pt]
\multicolumn{2}{@{}c@{\qquad}}{\hrulefill} & \multicolumn{2}{c@{\qquad}}{\hrulefill} & \multicolumn{2}{c@{}}{\hrulefill}\\
\textbf{Player} & $\bolds{\beta}$ & \textbf{Player} & $\bolds{\beta}$ & \textbf{Player} & $\bolds{\beta}$\\
\hline
\multicolumn{6}{@{}c@{}}{\textit{Basis} 5---\textit{frequency}}\\
G. Dragic & $-$1.286 & R. Foye & $-$1.325 & B. Bass & $-$1.378 \\
D. Lillard & $-$1.251 & C. Parsons & $-$1.306 & C. Frye & $-$1.357 \\
T. Burke & $-$1.183 & J. Anderson & $-$1.298 & S. Ibaka & $-$1.321 \\
W. Johnson & $-$1.163 & H. Barnes & $-$1.296 & C. Bosh & $-$1.312 \\
G. Hill & $-$1.121 & K. Korver & $-$1.282 & B. Griffin & $-$1.308
\\[3pt]
\emph{Average} & $-$1.031 & \emph{Average} & $-$1.184 & \emph{Average} & $-$1.325
\\[3pt]
S. Livingston & $-$0.911 & R. Allen & $-$1.097 & P. Millsap & $-$1.212 \\
M. Dellavedova & $-$0.903 & T. Hardaway Jr. & $-$1.079 & T. Thompson &
$-$1.190 \\
K. Walker & $-$0.894 & M. Barnes & $-$1.073 & Z. Randolph & $-$1.186 \\
D. Williams & $-$0.857 & I. Shumpert & $-$1.049 & T. Gibson & $-$1.159 \\
J. Jack & $-$0.819 & D. Waiters & $-$1.036 & T. Harris & $-$1.132\\
\hline
\end{tabular*}\vspace*{3pt}
\end{table}

As our main results we focus on parameters related to defensive shot selection
and shot efficiency effects. Here we focus on defensive results as
the novel contribution of this work, although offender-specific
parameters can be found in Supplement~A [\citet{frankssuppa}]. A sample
of the defensive logistic
regression log-odds for basis one (restricted area) and five
(center threes) are given in Tables~\ref{tabbasis1} and
\ref{tabbasis5}, respectively. For shot selection, we report the
defender effects,
$\beta_{jb}$, which correspond to the change in log-odds of a
shot occurring in a particular region, $b$, if defender $j$ guards
the offender for the entire possession. Smaller values correspond to
a reduction in the shooter's shot frequency in that region.

For shot efficiency we report $\phi_j+\xi_{b}\mathcal{D}^*_{jb}$, where
$\mathcal{D}^*_{jb}$ is player $j$'s difference in median distance
(relative to the average defender) to the offender
in region $b$. A~defender's overall effect on the outcome of a
shot depends on how close he tends to be to the shooter at the moment
the shot is taken, as well as the players' specific defensive skill
parameter $\phi_j$. Again, smaller values correspond to a reduction
in the shooter's shot efficiency, with negative values implying a
defender that is better than the global average.

%
\begin{figure}

\includegraphics{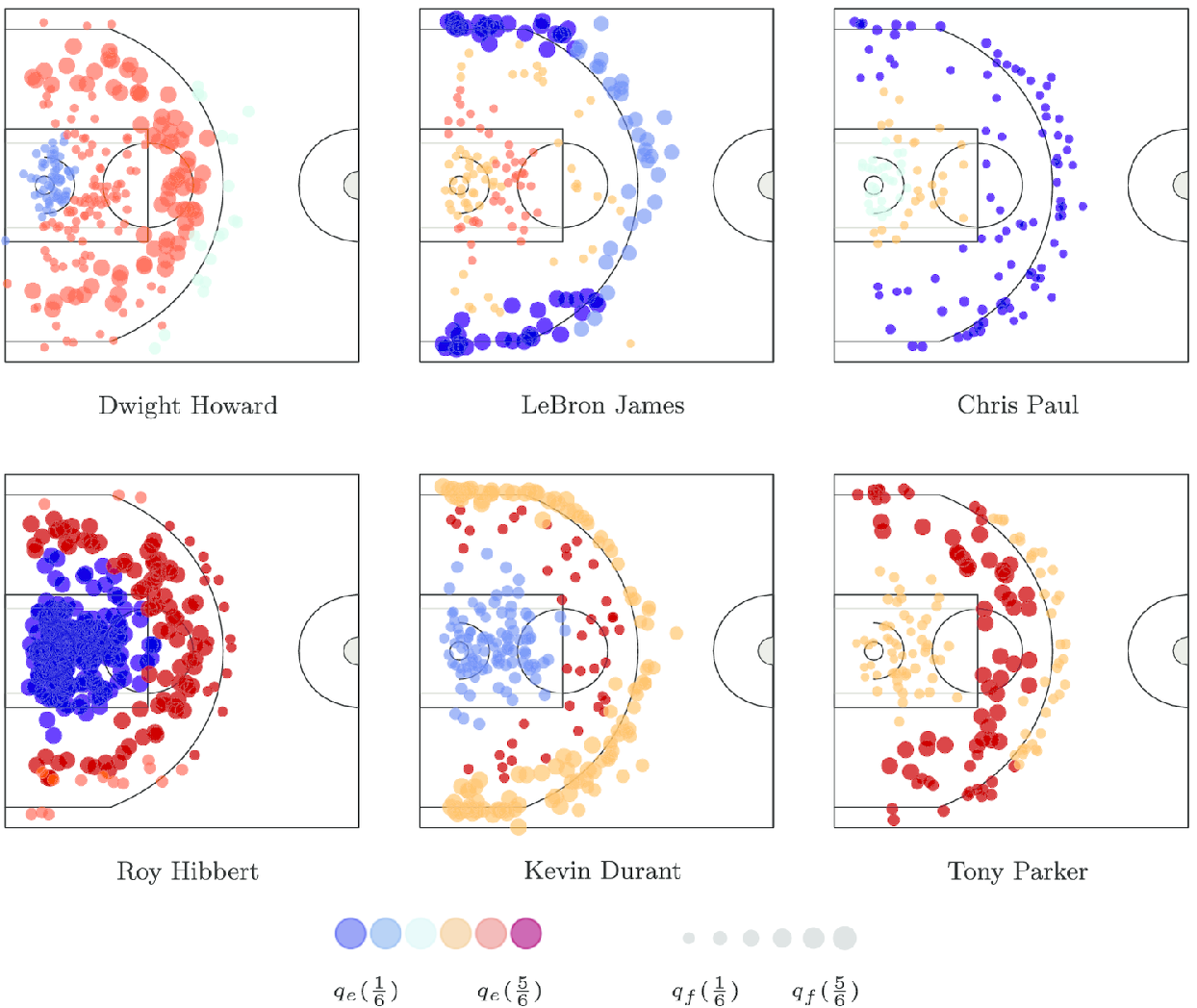}

\caption{Defensive shot charts. The dots represent the locations of
the shots faced by the defender, the color represents how the
defender changes the expected shot efficiency of shots, and the size
of the dot represents how the defender affects shot frequency, in
terms of the efficiency quantiles $q_e$ and frequency quantiles
$q_f$. Hibbert and Howard's contrasting defensive characteristics
are immediately evident. Small circles illustrate that, not
surprisingly, Chris Paul, the league leader in steals, reduces
opponents' shot frequency everywhere on the court.}
\label{figdshotchart}
\end{figure}

First, as a key point, we illustrate that defenders can affect shot
frequency (where an offender shoots) and shot efficiency (whether the
basket is made) and that, crucially, these represent distinct
characteristics of a defender. {This is well illustrated via two
well-regarded defensive centers, Dwight Howard and Roy Hibbert.
Roy Hibbert ranks first (Table~\ref{tabbasis1}) and fourth out of
167 defenders in his effect on shot efficiency in the paint\vadjust{\goodbreak} (bases 1
and 2). Dwight Howard, is ranked 50 and 117, respectively, out of 167
in these two bases. In shot selection, however, Dwight Howard ranks
11th and 2nd, respectively, in his suppression of shot attempts in the
paint (bases 1~and~2),\vadjust{\goodbreak} whereas Roy Hibbert ranks 161 in both bases 1
and 2. Whereas one defender may be good at discouraging shot
attempts, the other may be better at challenging shots once a
shooter decides to take it.}\vadjust{\goodbreak} 
{This demonstrates
that skilled defenders may impact the game in different ways, as a
result of team defensive strategy and individual skill. Figure~\ref
{figdshotchart} visually depicts the contrasting impacts of these
defenders.}

The defender effects do not always diverge so drastically between shot
efficiency and frequency, however. Some defenders are effective at
reducing both shot frequency and efficiency. For instance, Brandon Bass
is the top ranked defender in reducing both shot frequency and shot efficiency
in the perimeter (Table~\ref{tabbasis5}).

Importantly, our model is informative about how opposing shooters
perform against any defender in any region of the court. Even if a
defender rarely defends shots in a particular region, they may still
be partly responsible for giving up the shot in that region. As a point
guard, Chris Paul defends relatively few shots in basis 1, yet the
players he guards get fewer shots in this area relative to other
point guards (Table~\ref{tabbasis1}), perhaps in part because he
gets so many steals or is good at keeping players from driving toward
the rim. As a defender he spends very little time in this
court space, but we are still able to estimate how often his man beats
him to the basket for a shot attempt.

Finally, it is possible to use this model to help infer the best
defensive matchups. Specifically, we can infer the expected points
per possession a player should score if he were defended by a
particular defender. Fittingly, we found that one of the best defenders
on LeBron
James is Kawhi Leonard. Leonard received significant attention for
his tenacious defense on James in both the 2013 and 2014 NBA finals. Seemingly,
when the Heat play the Spurs and when James faces Leonard, we expect
James to score fewer points per possession than he would against almost
any other player.

While our results yield a detailed picture of individual defensive
characteristics, each defender's effect should only be interpreted in
the context of the team they play with. Certainly, many of these
players would not
come out as favorably if they did not play on some of the better defensive
teams in the league. For instance, how much a point guard reduces
opposing shot
attempts in the paint may depend largely on whether that defender
plays with an imposing center. Since basketball defense is inherently
a team sport, isolating true individual effects is likely not
possible without a comprehensive understanding of both team defensive
strategy and a model for the complex interactions between defenders.
Nevertheless, our model provides detailed summaries of individual
player effects in the context of their current team---a
useful measure in its own right. A full set of offender and defender
coefficients with standard errors can be found in  Supplement~A
[\citet{frankssuppa}].

%
%

\section{Discussion}

In this paper we have shown that by carefully constructing features
from optical player-tracking data, one is able to fill a current gap in
basketball analytics---defensive metrics. Specifically, our approach
allows us to characterize how players affect both shooting \textit
{frequency} and \textit{efficiency} of the player they are guarding.
By using an NMF-based decomposition of the court, we find an efficient
and data-driven characterization of common shot regions which naturally
corresponds to common basketball intuition. Additionally, we are able
to use this spatial decomposition to simply characterize the spatial
shot and shot-guarding tendencies of players, giving a natural
low-dimensional representation of a player's shot chart. Further, to
learn who is guarding whom,\vadjust{\goodbreak} we build a spatio-temporal model which is
fit with a combination of the EM-algorithm and generalized least
squares, giving simple closed-form updates for inference. Knowing who
is guarding whom allows for understanding of which players draw
significant attention, opening the court up for their teammates.
Further, we can see which teams induce a significant amount of
defensive switching, allowing us to characterize the ``chaos'' induced
by teams both offensively and defensively.

Combining this court representation and the mapping from offensive to
defensive players, we are able to learn how players inhibit (or
encourage) shot attempts in different regions of the court. Further,
conditioned on a shot being taken, we study how the defender changes
the probability of the shot being made. Moving forward, we plan to use
our results to understand the effects of coaching by exploring the
spatial characteristics and performance of players before and after
trades or coaching changes. Similarly, we intend to look at the
time-varying nature of defensive performance in an attempt to
understand how players mature in their defensive ability.

\section*{Acknowledgments}
The authors would like to thank STATS LLC
for providing us with the optical tracking data, as well as Ryan Adams,
Edo Airoldi, Dan Cervone, Alex D'Amour, Carl Morris and Natesh Pillai
for numerous valuable discussions.


\begin{supplement}[id=suppA]
  \sname{Supplement A}
  \stitle{Additional methods, figures and tables\\}
\slink[doi]{10.1214/14-AOAS799SUPPA} 
\sdatatype{.pdf}
\sfilename{aoas799\_suppa.pdf}
  \sdescription{We describe detailed methodology related to the shot
    type parameterizations and include additional graphics.  We also
    include tables ranking players' impact on shot frequency and
    efficiency (offense and defense) in all court regions.}
\end{supplement}

\begin{supplement}[id=suppB]
  \sname{Supplement B}
  \stitle{Animations}
\slink[doi]{10.1214/14-AOAS799SUPPB} 
\sdatatype{.zip}
\sfilename{aoas799\_suppb.zip}
  \sdescription{We provide GIF animations illustrating the ``who's
    guarding whom'' algorithm on different NBA possessions.}
\end{supplement}

%

\printaddresses

\begin{thebibliography}{25}
\bibitem[\protect\citeauthoryear{Bishop}{2006}]{bishop2006}
%
\begin{bbook}[mr]
\bauthor{\bsnm{Bishop},~\bfnm{Christopher~M.}\binits{C.~M.}}
(\byear{2006}).
\btitle{Pattern Recognition and Machine Learning}.
\bpublisher{Springer},
\blocation{New York}.
\bid{doi={10.1007/978-0-387-45528-0}, mr={2247587}}
\end{bbook}
%

\bptok{imsref}%
\endbibitem

\bibitem[\protect\citeauthoryear{B{\"o}hning}{1992}]{bohning92}
%
\begin{barticle}[mr]
\bauthor{\bsnm{B{\"o}hning},~\bfnm{Dankmar}\binits{D.}}
(\byear{1992}).
\btitle{Multinomial logistic regression algorithm}.
\bjournal{Ann. Inst. Statist. Math.}
\bvolume{44}
\bpages{197--200}.
\bid{doi={10.1007/BF00048682}, issn={0020-3157}, mr={1165584}}
\end{barticle}
%

\bptok{imsref}%
\endbibitem

\bibitem[\protect\citeauthoryear{Brunet et~al.}{2004}]{brunet}
%
\begin{barticle}[author]
\bauthor{\bsnm{Brunet},~\bfnm{Jean-Philippe}\binits{J.-P.}},
\bauthor{\bsnm{Tamayo},~\bfnm{Pablo}\binits{P.}},
\bauthor{\bsnm{Golub},~\bfnm{Todd~R.}\binits{T.~R.}} \AND
\bauthor{\bsnm{Mesirov},~\bfnm{Jill~P.}\binits{J.~P.}}
(\byear{2004}).
\btitle{Metagenes and molecular pattern discovery using matrix factorization}.
\bjournal{Proc. Natl. Acad. Sci. USA}
\bvolume{101.12}
\bpages{4164--9}.
\end{barticle}
%

\bptok{imsref}%
\endbibitem

\bibitem[\protect\citeauthoryear{Cervone et~al.}{2014}]{cervone}
%
\begin{bmisc}[author]
\bauthor{\bsnm{Cervone},~\bfnm{Dan}\binits{D.}},
\bauthor{\bsnm{D'Amour},~\bfnm{Alexander}\binits{A.}},
\bauthor{\bsnm{Bornn},~\bfnm{Luke}\binits{L.}} \AND
\bauthor{\bsnm{Goldsberry},~\bfnm{Kirk}\binits{K.}}
(\byear{2014}).
\btitle{POINTWISE: Predicting Points and Valuing Decisions in Real
Time with {NBA} Optical Tracking Data}.\vadjust{\goodbreak}
\end{bmisc}
%
\bptok{imsref}%
\endbibitem

\bibitem[\protect\citeauthoryear{Cressie}{1993}]{cressie93}
%
\begin{bbook}[mr]
\bauthor{\bsnm{Cressie},~\bfnm{Noel~A.~C.}\binits{N.~A.~C.}}
(\byear{1993}).
\btitle{Statistics for Spatial Data}.
\bpublisher{Wiley},
\blocation{New York}.
\bid{mr={1239641}}
\end{bbook}
%

\bptok{imsref}%
\endbibitem

\bibitem[\protect\citeauthoryear{Franks et~al.}{2015a}]{frankssuppa}
%
\begin{bmisc}[author]
{\bauthor{\bsnm{Franks},~\binits{A.}},
\bauthor{\bsnm{Miller},~\binits{A.}},
\bauthor{\bsnm{Bornn},~\binits{L.}} \AND
\bauthor{\bsnm{Goldsberry},~\binits{K.}}}
(\byear{2015a}).
\bhowpublished{Supplement to ``Characterizing the spatial structure of
defensive skill in professional basketball.''
DOI:\doiurl{10.1214/14-AOAS799SUPPA}}.
\bptok{imsref}%
\end{bmisc}
%

\bptok{imsref}%
\endbibitem

\bibitem[\protect\citeauthoryear{Franks et~al.}{2015b}]{frankssuppb}
%
\begin{bmisc}[author]
{\bauthor{\bsnm{Franks},~\binits{A.}},
\bauthor{\bsnm{Miller},~\binits{A.}},
\bauthor{\bsnm{Bornn},~\binits{L.}} \AND
\bauthor{\bsnm{Goldsberry},~\binits{K.}}}
(\byear{2015b}).
\bhowpublished{Supplement to ``Characterizing the spatial structure of
defensive skill in professional basketball.''
DOI:\doiurl{10.1214/14-AOAS799SUPPB}}.
\bptok{imsref}%
\end{bmisc}
%

\bptok{imsref}%
\endbibitem

\bibitem[\protect\citeauthoryear{Gelman and Rubin}{1992}]{gelman92}
%
\begin{barticle}[author]
\bauthor{\bsnm{Gelman},~\bfnm{Andrew}\binits{A.}} \AND
\bauthor{\bsnm{Rubin},~\bfnm{Donald~B.}\binits{D.~B.}}
(\byear{1992}).
\btitle{Inference from iterative simulation using multiple sequences}.
\bjournal{Statist. Sci.}
\bvolume{7}
\bpages{457--472}.
\end{barticle}
%

\bptok{imsref}%
\endbibitem

\bibitem[\protect\citeauthoryear{Goldsberry}{2012}]{goldsberry2012}
%
\begin{bmisc}[author]
\bauthor{\bsnm{Goldsberry},~\bfnm{Kirk}\binits{K.}}
(\byear{2012}).
\bhowpublished{Courtvision: New visual and spatial analytics for the NBA.
MIT Sloan Sports Analytics Conference.}
\end{bmisc}
%

\bptok{imsref}%
\endbibitem

\bibitem[\protect\citeauthoryear{Goldsberry}{2013}]{goldsberry2013}
%
\begin{bmisc}[author]
\bauthor{\bsnm{Goldsberry},~\bfnm{Kirk}\binits{K.}}
(\byear{2013}).
\bhowpublished{The Dwight Effect: A new ensemble of interior defense
analytics for the NBA.
MIT Sloan Sports Analytics Conference.}
\end{bmisc}
%

\bptok{imsref}%
\endbibitem

\bibitem[\protect\citeauthoryear{Kingman}{1992}]{kingman1992poisson}
%
\begin{bbook}[author]
\bauthor{\bsnm{Kingman},~\bfnm{John~Frank~Charles}\binits{J.~F.~C.}}
(\byear{1992}).
\btitle{Poisson Processes}.
\bpublisher{Oxford Univ. Press},
\blocation{London}.
\end{bbook}
%

\bptok{imsref}%
\endbibitem

\bibitem[\protect\citeauthoryear{Kubatko et~al.}{2007}]{kubatko}
%
\begin{barticle}[mr]
\bauthor{\bsnm{Kubatko},~\bfnm{Justin}\binits{J.}},
\bauthor{\bsnm{Oliver},~\bfnm{Dean}\binits{D.}},
\bauthor{\bsnm{Pelton},~\bfnm{Kevin}\binits{K.}} \AND
\bauthor{\bsnm{Rosenbaum},~\bfnm{Dan~T.}\binits{D.~T.}}
(\byear{2007}).
\btitle{A starting point for analyzing basketball statistics}.
\bjournal{J. Quant. Anal. Sports}
\bvolume{3}
\bpages{1--22}.
\bid{doi={10.2202/1559-0410.1070}, issn={1559-0410}, mr={2326663}}
\end{barticle}
%

\bptok{imsref}%
\endbibitem

\bibitem[\protect\citeauthoryear{Lee and Seung}{1999}]{lee1999learning}
%
\begin{barticle}[author]
\bauthor{\bsnm{Lee},~\bfnm{Daniel~D.}\binits{D.~D.}} \AND
\bauthor{\bsnm{Seung},~\bfnm{H.~Sebastian}\binits{H.~S.}}
(\byear{1999}).
\btitle{Learning the parts of objects by non-negative matrix factorization}.
\bjournal{Nature}
\bvolume{401}
\bpages{788--791}.
\end{barticle}
%

\bptok{imsref}%
\endbibitem

\bibitem[\protect\citeauthoryear{Lee and Seung}{2001}]{lee}
%
\begin{barticle}[author]
\bauthor{\bsnm{Lee},~\bfnm{Daniel~D.}\binits{D.~D.}} \AND
\bauthor{\bsnm{Seung},~\bfnm{H.~Sebastian}\binits{H.~S.}}
(\byear{2001}).
\btitle{Algorithms for non-negative matrix factorization}.
\bjournal{Adv. Neural Inf. Process. Syst.}
\bvolume{13}
\bpages{556--562}.
\end{barticle}
%

\bptok{imsref}%
\endbibitem

\bibitem[\protect\citeauthoryear{Limnios and Oprisan}{2001}]{semi2001}
%
\begin{bbook}[author]
\bauthor{\bsnm{Limnios},~\bfnm{Nikolaos}\binits{N.}} \AND
\bauthor{\bsnm{Oprisan},~\bfnm{Gheorghe}\binits{G.}}
(\byear{2001}).
\btitle{Semi-Markov Processes and Reliability}.
\bpublisher{Springer},
\blocation{Berlin}.
\end{bbook}
%

\bptok{imsref}%
\endbibitem

\bibitem[\protect\citeauthoryear{Macdonald}{2011}]{macdonald2011regression}
%
\begin{barticle}[author]
\bauthor{\bsnm{Macdonald},~\bfnm{Brian}\binits{B.}}
(\byear{2011}).
\btitle{A regression-based adjusted plus-minus statistic for NHL players}.
\bjournal{J.~Quant. Anal. Sports}
\bvolume{7}
\bpages{4}.
\end{barticle}
%

\bptok{imsref}%
\endbibitem

\bibitem[\protect\citeauthoryear{Maruotti and Ryd{\'e}n}{2009}]{Maruotti2008}
%
\begin{barticle}[mr]
\bauthor{\bsnm{Maruotti},~\bfnm{Antonello}\binits{A.}} \AND
\bauthor{\bsnm{Ryd{\'e}n},~\bfnm{Tobias}\binits{T.}}
(\byear{2009}).
\btitle{A semiparametric approach to hidden {M}arkov models under
longitudinal observations}.
\bjournal{Stat. Comput.}
\bvolume{19}
\bpages{381--393}.
\bid{doi={10.1007/s11222-008-9099-2}, issn={0960-3174}, mr={2565312}}
\bptnote{check year}%
\end{barticle}
%

\bptok{imsref}%
\endbibitem

\bibitem[\protect\citeauthoryear{Miller et~al.}{2014}]{miller2014}
%
\begin{binproceedings}[author]
\bauthor{\bsnm{Miller},~\bfnm{Andrew~C.}\binits{A.~C.}},
\bauthor{\bsnm{Bornn},~\bfnm{Luke}\binits{L.}},
\bauthor{\bsnm{Adams},~\bfnm{Ryan}\binits{R.}} \AND
\bauthor{\bsnm{Goldsberry},~\bfnm{Kirk}\binits{K.}}
(\byear{2014}).
\btitle{Factorized Point Process Intensities: A Spatial Analysis of
Professional Basketball}.
In \bbooktitle{Proceedings of the 31st International Conference on
Machine Learning (ICML)}.
\blocation{Beijing, China}.
\end{binproceedings}
%

\bptok{imsref}%
\endbibitem

\bibitem[\protect\citeauthoryear{M{\o}ller, Syversveen and
Waagepetersen}{1998}]{moller1998log}
%
\begin{barticle}[mr]
\bauthor{\bsnm{M{\o}ller},~\bfnm{Jesper}\binits{J.}},
\bauthor{\bsnm{Syversveen},~\bfnm{Anne~Randi}\binits{A.~R.}} \AND
\bauthor{\bsnm{Waagepetersen},~\bfnm{Rasmus~Plenge}\binits{R.~P.}}
(\byear{1998}).
\btitle{Log {G}aussian {C}ox processes}.
\bjournal{Scand. J. Stat.}
\bvolume{25}
\bpages{451--482}.
\bid{doi={10.1111/1467-9469.00115}, issn={0303-6898}, mr={1650019}}
\end{barticle}
%

\bptok{imsref}%
\endbibitem

\bibitem[\protect\citeauthoryear{Murphy}{2012}]{murphyml}
%
\begin{bbook}[author]
\bauthor{\bsnm{Murphy},~\bfnm{Kevin}\binits{K.}}
(\byear{2012}).
\btitle{Machine Learning: A Probabilistic Perspective}.
\bpublisher{MIT Press},
\blocation{Cambridge, MA}.
\end{bbook}
%

\bptok{imsref}%
\endbibitem

\bibitem[\protect\citeauthoryear{{N}ational {B}asketball {A}ssociation}{2014}]{nbaglossary}
%
\begin{bmisc}[author]
\bauthor{\bsnm{{N}ational {B}asketball {A}ssociation}}
(\byear{2014}).
\bhowpublished{A Glossary of NBA Terms.
Available at \url{http://www.NBA.com/analysis/00422966.html}.}
\end{bmisc}
%

\bptok{imsref}%
\endbibitem

\bibitem[\protect\citeauthoryear{Rosenbaum}{2004}]{rosenbaum}
%
\begin{bmisc}[author]
\bauthor{\bsnm{Rosenbaum},~\bfnm{Dan~T.}\binits{D.~T.}}
(\byear{2004}).
\bhowpublished{Measuring how NBA players help their teams win.
Available at \surl{82Games.com} (\url
{http://www.82games.com/comm30.htm}) 4--30.}
\end{bmisc}
%

\bptok{imsref}%
\endbibitem

\bibitem[\protect\citeauthoryear{Sill}{2010}]{sill2010improved}
%
\begin{binproceedings}[author]
\bauthor{\bsnm{Sill},~\bfnm{Joseph}\binits{J.}}
(\byear{2010}).
\btitle{Improved {NBA} adjusted plus-minus using regularization and
out-of-sample testing}.
In \bbooktitle{Proceedings of the 2010 MIT Sloan Sports Analytics Conference}.
\blocation{Boston, MA}.
\end{binproceedings}
%

\bptok{imsref}%
\endbibitem

\bibitem[\protect\citeauthoryear{{Stan Development Team}}{2014}]{stan}
%
\begin{bmisc}[author]
\bauthor{\bsnm{{Stan Development Team}}}
(\byear{2014}).
\bhowpublished{Stan: A C++ Library for Probability and Sampling,
Version~2.2}.
\end{bmisc}
%

\bptok{imsref}%
\endbibitem

\bibitem[\protect\citeauthoryear{Thomas et~al.}{2013}]{hockey}
%
\begin{barticle}[mr]
\bauthor{\bsnm{Thomas},~\bfnm{A.~C.}\binits{A.~C.}},
\bauthor{\bsnm{Ventura},~\bfnm{Samuel~L.}\binits{S.~L.}},
\bauthor{\bsnm{Jensen},~\bfnm{Shane~T.}\binits{S.~T.}} \AND
\bauthor{\bsnm{Ma},~\bfnm{Stephen}\binits{S.}}
(\byear{2013}).
\btitle{Competing process hazard function models for player ratings in
ice hockey}.
\bjournal{Ann. Appl. Stat.}
\bvolume{7}
\bpages{1497--1524}.
\bid{doi={10.1214/13-AOAS646}, issn={1932-6157}, mr={3127956}}\vadjust{\eject}
\end{barticle}
%
\bptok{imsref}%
\endbibitem

\bibitem[\protect\citeauthoryear{Yu}{2010}]{yu2010hidden}
%
\begin{barticle}[mr]
\bauthor{\bsnm{Yu},~\bfnm{Shun-Zheng}\binits{S.-Z.}}
(\byear{2010}).
\btitle{Hidden semi-{M}arkov models}.
\bjournal{Artificial Intelligence}
\bvolume{174}
\bpages{215--243}.
\bid{doi={10.1016/j.artint.2009.11.011}, issn={0004-3702}, mr={2724430}}
\end{barticle}
%

\bptok{imsref}%
\endbibitem
\end{thebibliography}
\end{document}